\documentclass[prb,amsmath,superscriptaddress,citeautoscript,twocolumn,
showpacs,floatfix]{revtex4}
\usepackage{bm}
\usepackage{amssymb}
\usepackage{graphicx}
\usepackage{amsmath, amsthm, amssymb, graphicx}

\usepackage{amsmath,amsthm,amsfonts,amscd,amssymb}
\usepackage{eucal} 	 	% Euler fonts
\usepackage{verbatim}      	% Allows quoting source with commands.
\usepackage{makeidx}       	% Package to make an index.
\usepackage{braket}

\usepackage{natbib}
\usepackage{bibentry}
\usepackage{hyperref}

\newcommand{\Oi}{\mathcal{O}}

\newcommand{\be}{\begin{equation}}
\newcommand{\ee}{\end{equation}}
\newcommand{\mb}{\mathbf}

\newcommand{\Li}{\mathcal{L}}

\begin{document}

\title{Structure of Correlated Worldline Theories of Quantum Gravity}

\author{A.O. Barvinsky}
\affiliation{Theory Department, Lebedev Physics Institute, Leninsky Prospect 53, Moscow 117924, Russia}

\author{D. Carney}
\affiliation{Pacific
Institute of Theoretical Physics, University of British Columbia,
6224 Agricultural Rd., Vancouver, B.C., Canada V6T 1Z1}
\affiliation{Joint Center for Quantum Information and
Computer Science, 3100 Atlantic Building, University of Maryland, College
Park, MD, USA 20742}
\affiliation{Joint Quantum Institute, National Institute of Standards and
Technology, 100 Bureau Dr., Gaithersburg, MD USA 20899}

\author{ P.C.E. Stamp}
 \affiliation{Pacific
Institute of Theoretical Physics, University of British Columbia,
6224 Agricultural Rd., Vancouver, B.C., Canada V6T 1Z1}
\affiliation{Department of Physics and Astronomy, University
of British Columbia, 6224 Agricultural Rd., Vancouver, B.C., Canada
V6T 1Z1}
\affiliation{School of Mathematics and Statistics, Victoria University of Wellington,
P.O. Box 600, Wellington 6140, New Zealand}

\date{{\small \today}}

\begin{abstract}
We consider the general form of "Correlated Worldline" (CWL) theories of quantum gravity. We show that one can have 2 different kinds of CWL theory, in which the generating functional is written as either a sum or a product over multiple copies of the coupled matter and gravitational fields. In both versions, the paths in a functional formulation are correlated via gravity itself, causing a breakdown of the superposition principle; however, the product form survives consistency tests not satisfied by the summed form. To better understand the structure of these two theories, we show how to perform diagrammatic expansions in the gravitational coupling for each version of CWL theory, using particle propagation and scalar fields as examples. We explicitly calculate contributions to 2-point and 4-point functions, again for each version of the theory, up to 2nd-order in the gravitational coupling. 

\end{abstract}

\pacs{03.65.Yz}

\maketitle

%%%%%%%%%%%%%%%%%%%%%%%%%%%%%%%%%%%%%%%%%%%%%%%%%%%%%%%%%%%%%
%%%%%%%%%%%%%%%%%%%%%%%%%%%%%%%%%%%%%%%%%%%%%%%%%%%%%%%%%%%%%

\section{ Introduction}
 \label{sec:intro}

%%%%%%%%%%%%%%%%%%%%%%%%%%%%%%%%%%%%%%%%%%%%%%%%%%%%%%%%%%%%%
%%%%%%%%%%%%%%%%%%%%%%%%%%%%%%%%%%%%%%%%%%%%%%%%%%%%%%%%%%%%%

Over the last few decades various attempts have been made to find a quantum theory of gravity which not only deals with the usual field-theoretic issues (renormalizability, etc.), but which also addresses some of the perceived inconsistencies between classical relativity and quantum field theory \cite{ashtekar74,carlip01,penrose96,wald84,amps13}. These issues become particularly acute when considering black holes, and the infamous ``black hole information paradox", which in its original formulation \cite{hawkingBH} indicated that black holes must cause a breakdown of quantum mechanics, because information would disappear behind the event horizon. There are at present many points of view on this paradox \cite{banks94,braunstein07,unruh12}, and the problem is currently under intense discussion.

However, the idea that gravitation might lead to a breakdown of quantum theory is actually rather old, and has been connected in various ways with the interpretational problems of quantum mechanics \cite{karolhazy,kibbleb,page81,unruh84,penrose96,stamp12,stamp15}. Some of the key issues are as follows:

(i) If one assumes that the metric field $g^{\mu\nu}(x)$ must be quantized like any other field, then superpositions of matter fields automatically involve superpositions of different $g^{\mu\nu}(x)$. However, amongst other things $g^{\mu\nu}(x)$ defines causal relations between events, so that such superpositions, or more generally quantum fluctuations of $g^{\mu\nu}(x)$, means that we lose our usual notions of causality, which are a fundamental part of conventional quantum field theory (QFT). If on the other hand we treat $g^{\mu\nu}(x)$ classically, but all other fields quantum mechanically, the result is apparently inconsistent \cite{page81,unruh84,duff82,eppleyH77}. It is thus usually assumed that $g^{\mu\nu}(x)$ must also be quantized, although there are dissenting points of view \cite{mattingly06,carlip08}, and suggestions exist for how to test this question experimentally \cite{vedral17}.

(ii) How, in any quantum theory of gravity, are we supposed to handle the classical regime? Do we simply impose a ``von Neumann cut" between quantum and classical worlds, with all the usual associated problems (which include the inconsistency of ``wave function collapse" with basic conservation laws), or try and quantize everything, and then somehow derive the classical world from a totally quantized treatment (see, eg., ref. \cite{hartleGM}), as a limiting case?

It is questions like these, along with general unease about quantum mechanics at the macroscopic level, which have led to the suggestions that gravity may cause a breakdown of quantum mechanics. Physical arguments in this direction, using thought experiments involving mass superpositions, have been given by several authors \cite{penrose96,kibbleb,page81,unruh84,stamp12,stamp15}. There have also been more concrete suggestions for new kinds of theory. The first real attempt was made by Kibble and co-workers \cite{kibbleb,kibble1,kibble2}, who introduced non-linear terms into the basic dynamics of quantum mechanics and quantum field theory (immediately implying a breakdown of the superposition principle); the source of this non-linearity was sought in gravity. By coupling $g^{\mu\nu}(x)$ to a $\langle \psi | T_{\mu\nu}(x) | \psi \rangle$ involving a matter state $|\psi\rangle$, one immediately obtained a violation of the superposition principle, because of the non-linear dynamics of $g^{\mu\nu}(x)$. This was the first clear attempt to derive a breakdown of linearity in the Schrodinger equation (or its field-theoretical generalizations) from the intrinsic non-linearity of General Relativity.

There were two problems with this approach, both immediately identified by Kibble. The first was similar to the ``classical/quantum" mixing noted above - coupling a quantum metric field to a $c$-number matter field expectation value leads to paradoxes (also emphasized by Unruh \cite{unruh84}). The second problem came from the difficulty of grafting the usual quantum-mechanical ideas about measurements, states, and operators onto non-linear modifications of quantum theory \cite{kibble1}. An attempt to get past problems like this was made by Weinberg \cite{weinberg79}; unfortunately other difficulties then arose, notably superluminal signal propagation and non-violation of Bell inequalities \cite{gisin80,polchinski80}.

Later discussions by Penrose argued that any attempt to create gravitational superpositions would cause ``gravitational decoherence" between different branches of the superposition. He attempted to quantify this, initially using arguments in the Newtonian limit \cite{penrose96}, and then using a using a Schrodinger-Newton equation \cite{penrose98}. Similar ideas have been discussed, also using some kind of Schrodinger-Newton approach, in work by other authors \cite{diosi,bassi,adler}. A specific experiment to test ideas of this kind, and differentiate them from the predictions of any quantum theory, was then proposed in 2003 by Penrose, Bouwmeester, and co-workers \cite{bouwm03}. 

More recently one of us has made an attempt at a field theory which tries to get past these difficulties \cite{stamp12,stamp15}, called ``correlated worldline" (CWL) theory .  In a CWL theory, both matter and gravitational fields are quantized, but we postulate that (i) correlations appear between different paths for these fields - causing a breakdown of the superposition principle - and (ii) that the correlations are gravitational - the metric couples different paths to each other, whether they be paths of the spacetime metric itself, or simply matter field paths. The  formulation of CWL theory described in these papers \cite{stamp15} was given in what we will call the ``summation" form, explained immediately below.

A key question in any work of this kind is whether or not the theory is internally consistent. To look at the consistency of CWL theory is the first of the two main goals of the present paper. Consistency checks include generalized Ward and Noether identities, and checks on the classical limit. All of these will be discussed below.  

In the course of trying to establish such consistency, we have found that the choice of a CWL field theory is not unique; apart from the summation form discussed already \cite{stamp12,stamp15}, one can also define a class of ``product CWL" theories. As far as we can tell these two classes of CWL theory exhaust the possible theories of this type.

Roughly speaking the difference between the summed and product forms of CWL theory is as follows: the summation form writes, for a field $\phi(x)$ coupled to a current $J(x)$, the generating functional
\begin{equation}
\mathbb{Q}[J] \;=\;  \oint {\cal D}g \; e^{{i \over \hbar} S_G[g]} \sum_{n=1}^{\infty} {1 \over n!}\;Q_n[g,J]  \;\;\;\;\; (summed) \;\;\;
 \label{Q-sch1}
\end{equation}
in which $Q_n[g,J]$ is that contribution to the sum coming from an $n$-tuple of paths; whereas the product version writes
    \begin{eqnarray}
    &&\mathbb{Q}[J]=\prod\limits_{n=1}^\infty \tilde{Q}_n[\,J\,]  \;\;\;\;\;\; (product) \;\;\; \\
    &&\tilde{Q}_n[\,J\,]=
    \oint {\cal D}g \; e^{{i \over \hbar} S_G[g]} \,\left({\cal Z}[g,
    J/c_n]\right)^n     \label{bbQ-J01}
    \end{eqnarray}
where ${\cal Z}$ is just the particle generating functional in conventional field theory;  we are now summing over logarithms, and $\ln \mathbb{Q}[J]$ -- the generator of connected graphs -- is just a sum of contributions of single $g$ integrals. There are regulators in both the $Q_n[g,J]$ in (\ref{Q-sch1}) and  $\tilde{Q}_n[J]$ in (\ref{bbQ-J01}), which will be discussed when we develop the formal details (the regulator in $\tilde{Q}_n[J]$ is denoted by $c_n$). In (\ref{Q-sch1}) and (\ref{bbQ-J01}) we have also suppressed Faddeev-Popov gauge fixing factors - these will be reinstated in the formal development below.

The generating functional $\mathbb{Q}[J]$ generates connected correlation functions, of the usual form:
\begin{equation}
{\cal G}_{\ell}(\{ x_k \}) \;=\; \left( {\hbar \over i} \right)^{\ell} \lim_{J(x) \rightarrow 0} \left[{ \delta^{\ell} \ln \mathbb{Q}[J] \over \delta J(x_1 ) .. \delta J(x_{\ell})}\right]
 \label{Gn-corr}
\end{equation}
no matter whether we deal with the summed CWL or product CWL theories.

As noted above, a key goal of the present paper is to find out whether such theories are internally consistent, a first step in establishing their viability as descriptions of Nature. The second main goal is to show how perturbative expansions (in the inverse Planck mass) may be carried out. There are 2 main reasons for this. First, these calculations are essential for any application of the theory to laboratory experiments; and second, their viability also constitutes a check on the theory.

Of course another way to decide on the viability of a theory is to compare its predictions with experiment. Initial calculations with CWL theory \cite{stamp12,stamp15} predicted a crossover to classical dynamics, for both the matter and gravitational fields, when masses are sufficiently large (of order the Planck mass $m_p$, although the actual crossover value depends quite strongly on the detailed structure of the bodies involved). This crossover to the classical regime is caused by the gravitational correlations, and does not appear to involve decoherence - it can be characterized as a ``path bunching" effect, whereby the paths of massive objects are forced to stay close to each other because of their strong correlations. These prediction are very different from those of Penrose's theory \cite{bouwm03,bouwm08}.  

At the present time no experiment has actually seen any deviations from quantum mechanics, and the experiments are still at an early stage. Thus at the present time our best recourse is to explore possible new theories, in the way being done here.

In what follows we will examine both the sum and product forms of CWL theory. We begin in the next section with the summation form \cite{stamp12,stamp15}, and show that this form satisfies the relevant Noether identities, and has a well-defined perturbation expansion, which is exhibited in section \ref{sec:pertSum} by calculating propagators for the simple examples of particle dynamics and a 2-dimensional scalar field. Then, in section \ref{sec:cons}, we note that the summation version of CWL theory has a peculiar classical limit, in which quantum fluctuations disappear in a discontinuous way as $\hbar \rightarrow 0$. We then introduce the ``product" version of CWL theory, in section \ref{sec:CWLprod}; this turns out to have a perfectly continuous classical limit, and we derive its basic properties. After doing this, we move on in section \ref{sec:pertProd} to develop perturbative expansions in the gravitational coupling for this product theory, exhibiting their diagrammatic form. Finally, in the last section we summarize our conclusions from these investigations.

%%%%%%%%%%%%%%%%%%%%%%%%%%%%%%%%%%%%%%%%%%%%%%%%%%%%%%%%%%%%%
%%%%%%%%%%%%%%%%%%%%%%%%%%%%%%%%%%%%%%%%%%%%%%%%%%%%%%%%%%%%%

\section{CWL Theory: Summation Version}
 \label{sec:CWLsum}

%%%%%%%%%%%%%%%%%%%%%%%%%%%%%%%%%%%%%%%%%%%%%%%%%%%%%%%%%%%%%
%%%%%%%%%%%%%%%%%%%%%%%%%%%%%%%%%%%%%%%%%%%%%%%%%%%%%%%%%%%%%

We briefly recall here the basic formal structure of the summation version of the CWL theory \cite{stamp15}; the main purpose here is to set up the perturbation expansion of the next section, and the consistency analysis of section \ref{sec:cons}, and to establish notation. In what follows we compare the CWL structure with that of conventional field theory.

%%%%%%%%%%%%%%%%%%%%%%%%%%%%%%%%%%%%%%%%%%%%%%%%%%%%%%%%%%%%%
\subsection{Generating Functional}
 \label{sec:GenF-sum}
%%%%%%%%%%%%%%%%%%%%%%%%%%%%%%%%%%%%%%%%%%%%%%%%%%%%%%%%%%%%%

In the summation version of CWL theory one implements the idea of correlating paths \cite{stamp12,stamp15} by starting from a generating functional written in the form given in (\ref{Q-sch1}). The easiest way to see how this works is via examples.

\vspace{2mm}

{\it Example 1: Massive particle}: Consider a single massive relativistic particle; in a given fixed background metric $g^{\mu \nu}$ this has the action
\begin{equation}
S_o[g,q] = m \int d^4x d\tau \delta(x - q(\tau)) \sqrt{-g_{\mu \nu} \dot{q}^{\mu} \dot{q}^{\nu}}
  \label{QMpart}
\end{equation}
where $\dot{q} \equiv dq/d\tau$ denotes a derivative with respect to proper time, and $g^{\mu\nu}(x)$ is the spacetime metric.

\begin{figure}
\includegraphics[width=3.2in]{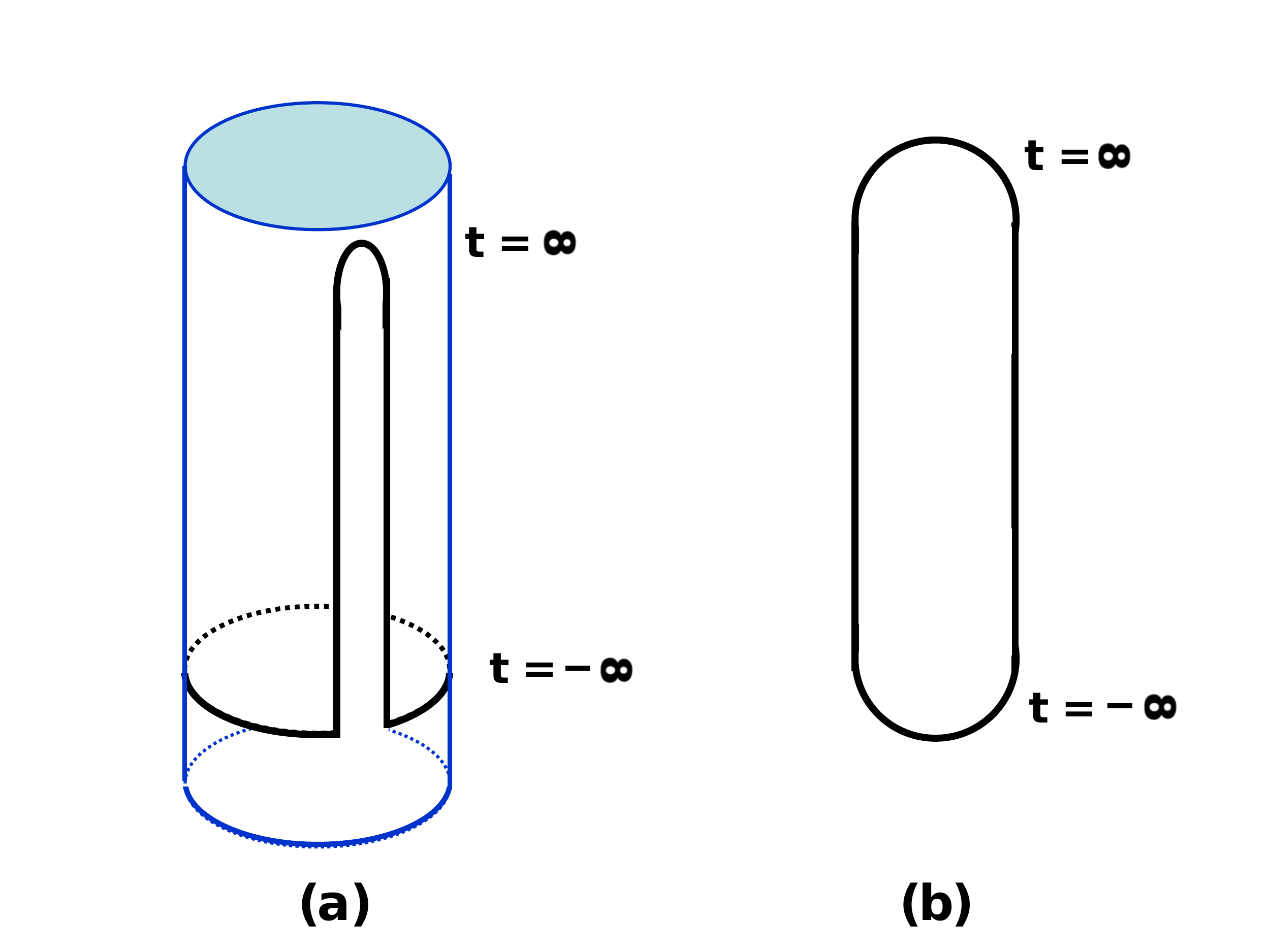}
\caption{\label{fig:ring} The contour involved in the ``ring" diagrams for the generating functional in this paper. In (a) we show in bold black the contour for a matter particle or field in proper time, extending from $t = -\infty$ up to $t = \infty$ and back again; it is then closed at $t = -\infty$ around the ``temperature cylinder" of circumference $2\pi/kT$. In (b) we show how such a propagator is represented in this paper. }
\end{figure}

In conventional field theory one then has a generating functional for the coupled dynamics of the particle and the metric given by
\begin{eqnarray}
{\cal Z}[j] &=& \oint {\cal D}g^{\mu\nu} \Delta(g) \; e^{{i \over \hbar} S_G[g]}\oint {\cal D}q \; e^{{i \over \hbar} \left(S_o[q,g] + \int jq \right)   } \nonumber \\
& \equiv & \oint {\cal D}g^{\mu\nu} \Delta(g) \; e^{{i \over \hbar} S_G[g]} {\cal Z}[g,j]
 \label{genF-p}
\end{eqnarray}
where $\oint {\cal D}g^{\mu \nu}$ is the functional integral over the metric, with a Faddeev-Popov determinant $\Delta(g)$ which divides out diffeomorphism-equivalent metric configurations, and $j(\tau)$ is an external field coupling linearly to $q(\tau)$. In this paper we will assume ``closed path" or ``ring" functional integration, denoted by $\oint$, in which one proceeds from a time slice at past infinity, out to future infinity, and back again \cite{SdW-GF,schmid82}. In conventional Schwinger-Keldysh theory the particle would be injected at past infinity on the ``in" line, and then recovered at past infinity on the ``out" line. Here we will assume thermal equilibrium, and close the path at past infinity, with the ring integration \cite{bloch} completed along an imaginary time contour around a cylinder of circumference $1/kT$ (see Fig. \ref{fig:ring} ).

%%%%%%%%%%%%%%%%%%%%%%%%%%%%%

%%%%%%%%%%%%%%%%%%%%%%%%%%%%%

In summed CWL theory, on the other hand, we have a generating functional for the coupled metric/particle dynamics taking the form given in (\ref{Q-sch1}). Written out in full this then gives
\begin{widetext}
\begin{equation}
\mathbb{Q}[j] \;=\;  \oint {\cal D}g^{\mu \nu} \Delta(g) \; e^{{i \over \hbar} S_G[g^{\mu \nu}]} \sum_{n=1}^{\infty} {1 \over n!} \prod_{k=1}^n \oint {\cal D}q_k \; e^{{i \over n \hbar } \sum_k \;   (S_o[q_k, g^{\mu \nu}] + \int dt j(t) q_k(t))}
 \label{K2-CWL}
\end{equation}

\end{widetext}
so that, by comparing this result with (\ref{Q-sch1}), we see that the factor $Q_n[g,J]$ in (\ref{Q-sch1}) is just
\begin{equation}
Q_n[g,J] \;=\; \prod_{k=1}^n \oint {\cal D}q_k \; e^{{i \over n \hbar } \sum_k \;   (S_o[q_k, g^{\mu \nu}] + \int dt j(t) q_k(t))}
 \label{Q-n-sum})
\end{equation}

Note the factor of $1/n$ in the exponential, multiplying both the action and the current; this is a regulator, discussed in detail in ref. \cite{stamp15}, to which we return below.

\vspace{3mm}

{\it Example 2: Massive scalar field}: Consider now a scalar field $\phi(x)$ with action
\be
S_M =  - \frac{1}{2} \int d^4 x \sqrt{-g} \; (g^{\mu \nu} \partial_{\mu} \phi \partial_{\nu} \phi - 2 V(\phi))
\ee
where $V(\phi)$ is some local interaction, and the theory is assumed renormalizable and stable.

%%%%%%%%%%%%%%%%%%%%%%%%%%%%%

\begin{figure}
\includegraphics[width=3.2in]{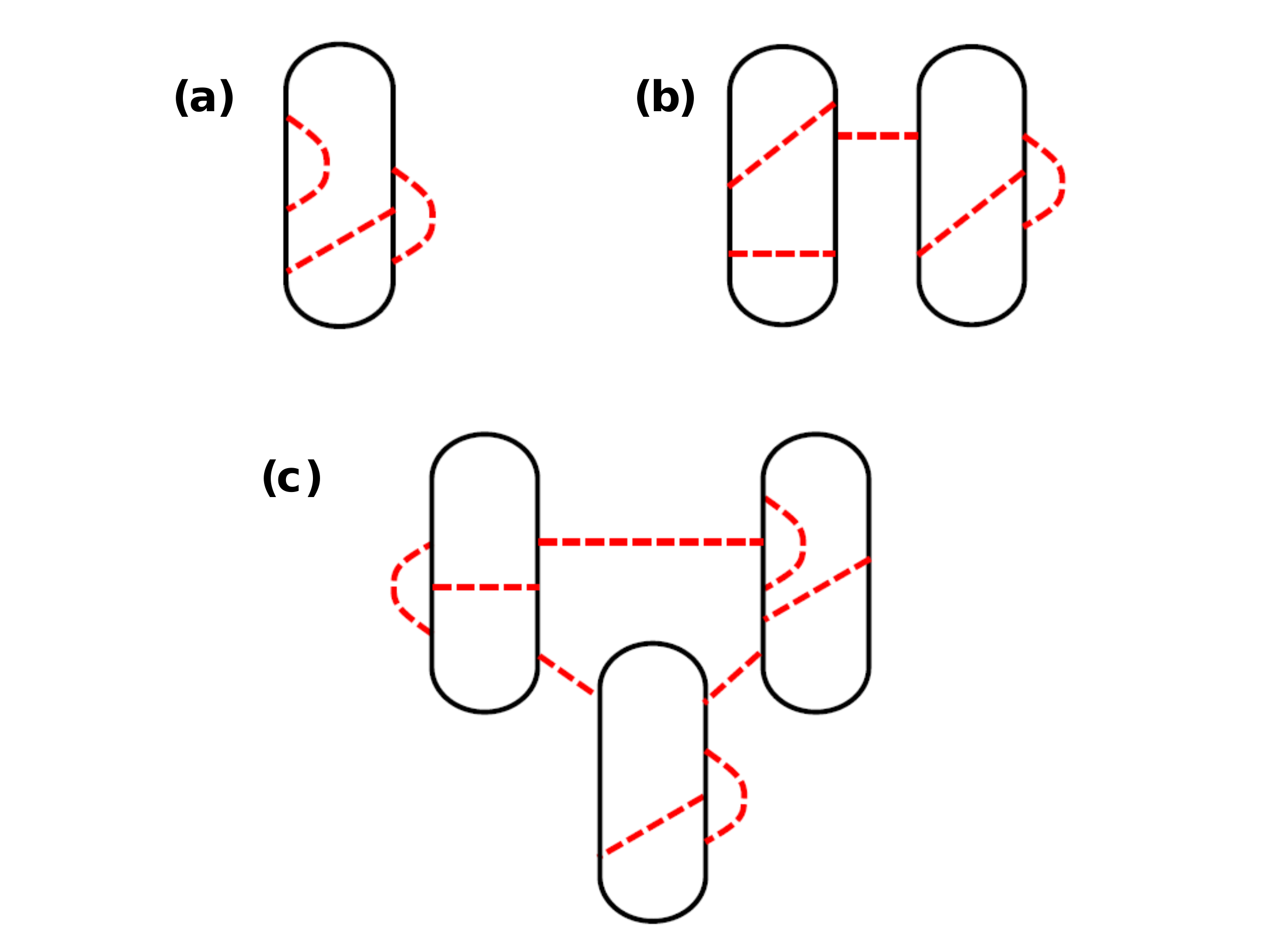}
\caption{\label{fig:CWLgen} Feynman-Keldysh diagrams contributing to the generating functional $\mathbb{Q}[J]$ for summed CWL theory, when we set the external current $J(x) = 0$; compare eqtns. (\ref{phi-CWL}) and (\ref{Qn-gJ}). In (a) we see a term contributing to $Q_1[J]$; in (b) a term contributing to $Q_2[J]$; and in (c) a term contributing to $Q_3[J]$. The scalar field propagator is shown in black, and the graviton propagator in red. }
\end{figure}

%%%%%%%%%%%%%%%%%%%%%%%%%%%%%

Then, this theory has in CWL the generating functional (see Fig. \ref{fig:CWLgen}):
\begin{eqnarray}
\mathbb{Q}[J] \;&=&\;  \oint {\cal D}g  \; e^{{i \over \hbar} S_G[g]} \sum_n {1 \over n!} \; Q_n[J,g] \nonumber \\
&\equiv &\; \sum_n {1 \over n!} \; \tilde{Q}_n[J]
  \label{phi-CWL}
\end{eqnarray}
where the external current $J(x)$ couples to $\phi(x)$, where
\begin{equation}
Q_n[J,g] = \prod_{k=1}^n \oint {\cal D}\phi_k \; e^{{i \over n \hbar } \sum_k \;   (S_M[\phi_k, g^{\mu \nu}] + \int J \phi_k)}
 \label{Qn-gJ}
\end{equation}
and where $\tilde{Q}_n[J] = \oint {\cal D} g^{\mu\nu} \Delta(g) e^{{i \over \hbar} S_G[g]} Q_n[J,g]$ is the result of integrating out the gravitational modes from $Q_n[J,g]$. Note that $Q_1[J]$ is just the same as ${\cal Z}[J]$, the generating functional for standard quantum gravity.

We can also write generating functionals for some general set of functions of the fields. Thus consider some field $\psi(x)$ (fermionic or bosonic), and define a set of functions
\be
\Oi_a(x_a) = \Oi_a(\psi(x_a),\partial_{\mu}\psi(x_a),\ldots), \ \ \ a = 1,\ldots,N
 \label{genF-O}
\ee
which are local functions of the field $\psi$ and its derivatives, evaluated at a specific spacetime event.  We then write a generating functional wherein each function $\Oi_a$ is coupled to an external current $J_a$ which couples to $\Oi_a(\psi, \partial \psi)$, so that
\begin{widetext}
\begin{equation}
Q_n[\{ J_a \},g] = \prod_{k=1}^n \oint {\cal D}\psi_k \; \exp {i \over n \hbar } \sum_k \;   (S_M[\psi_k, g^{\mu \nu}] + \int \sum_a J_a(x) \Oi_a(\psi_k(x))
 \label{Qn-gJa}
\end{equation}
\end{widetext}

%%%%%%%%%%%%%%%%%%%%%%%%%%%%%%%%%%%%%%%%%%%%%%%%%%%%%%%%%%%%%
\subsection{Correlation functions and Propagators}
 \label{sec:G+K}
%%%%%%%%%%%%%%%%%%%%%%%%%%%%%%%%%%%%%%%%%%%%%%%%%%%%%%%%%%%%%

In any CWL theory it is important to distinguish between propagators and correlators. In what follows we will define ordinary propagators for particles and fields, and discuss how these are distinguished from correlation functions for the same fields, in the summed CWL theory.

%%%%%%%%%%%%%%%%%%%%%%%%%%%%%%%%%%%%%%%%%%%%%%%
\subsubsection{Propagators}
 \label{sec:prop}
%%%%%%%%%%%%%%%%%%%%%%%%%%%%%%%%%%%%%%%%%%%%%%%

In ordinary quantum field theory or relativistic particle mechanics, one defines the propagator in a way directly related to the generating functional. Thus  for a massive relativistic particle coupled to the background dynamic metric, one defines a particle propagator
\begin{equation}
K(x,x'|j) \;=\; \oint {\cal D}g \; e^{{i \over \hbar} S_G[g]}\int_{x'}^x {\cal D}q \; e^{{i \over \hbar} \left(S_o[q,g] + \int jq \right)   }
 \label{GF-p}
\end{equation}
between spacetime points $x'$ and $x$ in the presence of a perturbing field $j(\tau)$. Here again we assume a ring functional integration over a metric in equilibrium at temperature $T$, and we employ the shorthand
\be
\int_{x'}^{x} Dq_k \;\equiv\; \int_{q(\tau_i) = x'}^{q(\tau_f) = x} Dq_k
 \label{PI-def}
\ee
for the path integration limits, where $\tau_{i}$ and $\tau_f$ are the initial and final proper times for all paths $q(\tau)$ beginning at spacetime point $x'$ and ending at spacetime point $x$.  Setting $j(\tau) = 0$ gives us back the usual form for the propagator $K(x,x')$.

In summed CWL theory the propagator ${\cal K}(x,x'|j)$ for the particle is related to $\mathbb{Q}[j]$ in the same way; we have, written out in full, the result
\begin{widetext}
\begin{equation}
{\cal K}(x,x'|j) \;=\;  \int {\cal D}g^{\mu \nu} \Delta(g) \; e^{{i \over \hbar} S_G[g^{\mu \nu}]} \sum_{n=1}^{\infty} {1 \over n!} \prod_{k=1}^n \int_{x'}^x {\cal D}q_k \; e^{{i \over n \hbar} \sum_k \;  (S_0[q_k, g^{\mu \nu}] + \int dt j(t) q_k(t))}
 \label{K2-CWL}
\end{equation}
with an obvious generalization to field propagators; thus,
the amplitude to propagate between two configurations $\Phi'(x)$ and $\Phi_(x)$ of a scalar field $\phi(x)$ in the CWL theory is just:
\begin{equation}
{\cal K}(\Phi,\Phi'|J)\;=\;  \oint {\cal D}g^{\mu \nu} \Delta(g) \; e^{{i \over \hbar} S_G[g^{\mu \nu}]} \; \sum_{n=1}^{\infty} {1 \over n!} \prod_{k=1}^n \int^{\Phi}_{\Phi'} {\cal D}\phi_k \; e^{{i \over n \hbar } \sum_k \;   (S_M[\phi_k, g^{\mu \nu}] + \int dt J(x) \phi_k(x))}
 \label{Kphi-CWL}
\end{equation}

\end{widetext}

Again, setting the external currents $j(\tau)$ or $J(x)$ to zero gives back conventional propagators ${\cal K}(x,x')$ and ${\cal K}(\Phi,\Phi')$ respectively. These propagators can be represented diagramatically \cite{stamp15}.

%%%%%%%%%%%%%%%%%%%%%%%%%%%%%%%%%%%%%%%%%%%%%%%
\subsubsection{Correlation Functions}
 \label{sec:corr}
%%%%%%%%%%%%%%%%%%%%%%%%%%%%%%%%%%%%%%%%%%%%%%%

Correlation functions are defined in CWL theory just as in standard QFT, starting from the the CWL generating functional $\mathbb{Q}[j]$. Thus, for our relativistic particle coupled to the metric field, we define connected correlation functions $G_n^{\sigma_1,..\sigma_n}(s_1,..s_n)$ at proper times $s_k$, with $k = 1,2,..n$, and where the index $\sigma_k = \pm$ indicates upon which section of the loop (forward or backward) the functional differential is being taken. Then, in the usual way, we write $G_n^{\sigma_1,..\sigma_n}(s_1,..s_n)$ in terms of a set of functional differentials of $\mathbb{Q}[j]$ as 
\begin{equation}
G_n^{\{ \sigma_k \}}(\{ s_k \}) \;=\;    (\hbar/i)^n \left[ { \delta^n \ln \mathbb{Q}[j] \over \delta j^{\sigma_1}(s_1) .. \delta j^{\sigma_n}(s_n)} \right]\Bigg|^{j=0} \;\;\;
 \label{Gn-NR}
\end{equation}
where we have written $G_n^{\sigma_1,..\sigma_n}(s_1,..s_n)$ in abbreviated form as $G_n^{\{ \sigma_k \}}(\{ s_k \})$, and where
$\delta j^{\sigma_k}(s_k)$ is the external current insertion at proper time $s_k$, on either the forward/backward section of the loop (for which $\sigma_k = \pm 1$ respectively).

\begin{figure}
\includegraphics[width=3.2in]{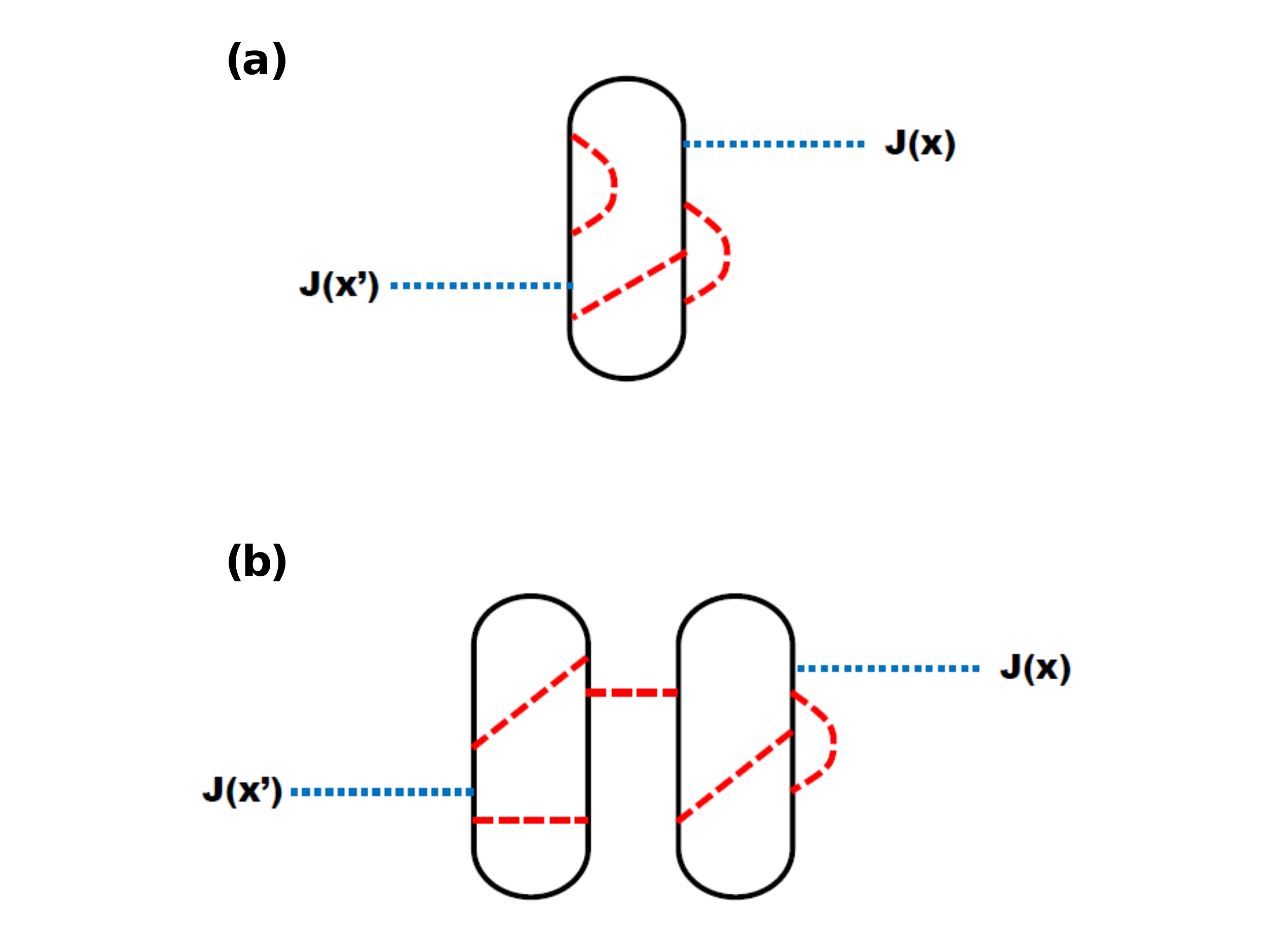}
\caption{\label{fig:CWLcorr} Typical diagrams contributing to a correlator in summed CWL theory (cf. eqtn. (\ref{Gsig-n})). Here we show graphs for $G_2^{+-}(x,x')$, a 2-point correlator in which one current insertion is made on the forward ($+$) path, and one on the backward ($-$) path. Contributions come from al of the $\tilde{Q}_n[J]$ in $\mathbb{Q}[J]$.  In (a) we see a contribution from $Q_1[J]$; in (b) a contribution from $Q_2[J]$. The scalar field propagator is shown in black, the graviton propagator in red, and current insertions in blue. }
\end{figure}

For fields the procedure is a trivial generalization of this - thus, eg., the simple scalar field system discussed above
has connected correlators of form: 
\begin{equation}
G_n^{\{ \sigma_k \}}(\{ x_k \}) \;=\;    (\hbar/i)^n \left[ { \delta^n \ln \mathbb{Q}[J] \over \delta J^{\sigma_1}(x_1) .. \delta J^{\sigma_n}(x_n)} \right]\Bigg|^{J=0} \;\;\;
 \label{Gsig-n}
\end{equation}
where we now have external current insertions $\delta J^{\sigma_k}(x_k)$ at spacetime points $\{ x_k \}$. Again, a diagrammatic representation of this equation can be given \cite{stamp15} (compare Fig. \ref{fig:CWLcorr}).

%%%%%%%%%%%%%%%%%%%%%%%%%%%%%

%%%%%%%%%%%%%%%%%%%%%%%%%%%%%

From now on, to reduce clutter in the equations, we will drop reference to the path indices $\{ \sigma_k \}$; they can be reinstated with no difficulty.

The expectation value of general time-ordered products of local field functions (compare eq. (\ref{genF-O})) can be calculated in a similar way. Suppose we want to calculate something like
\be
\label{generalopcorr}
F(x_1,\ldots,x_n) = \braket{ 0 | T \Oi_1(x_1) \cdots \Oi_N(x_n) | 0 }
\ee
where, again, we drop the path indices $\{ \sigma_k \}$.
Then \eqref{generalopcorr} is just given by 
\be
F(x_1,\ldots,x_n) = (\hbar/i)^n) \frac{\delta^{n} \ln \mathbb{Q}[\{J_a \}]}{\delta J_1(x_1) \cdots \delta J_N(x_n)} \Bigg|^{J=0}.
 \label{FxQJ}
\ee
This reproduces, in particular, the correct results for the simple case where the operators $\Oi$ are just the fields $\psi$ themselves.

%%%%%%%%%%%%%%%%%%%%%%%%%%%%%%%%%%%%%%%%%%%%%%%%%%%%%%%%%%%%%%%%%%%%%%%%%%%%
%%%%%%%%%%%%%%%%%%%%%%%%%%%%%%%%%%%%%%%%%%%%%%%%%%%%%%%%%%%%%%%%%%%%%%%%%%%%

\section{Perturbation expansions for Summed CWL theory}
\label{sec:pertSum}

%%%%%%%%%%%%%%%%%%%%%%%%%%%%%%%%%%%%%%%%%%%%%%%%%%%%%%%%%%%%%%%%%%%%%%%%%%%%
%%%%%%%%%%%%%%%%%%%%%%%%%%%%%%%%%%%%%%%%%%%%%%%%%%%%%%%%%%%%%%%%%%%%%%%%%%%%

A simple way to probe the structure of a class of theories is to see how one calculates physical quantities in perturbation theory. In this section we do low-order perturbative expansions in the gravitational coupling for propagators in the summation version of CWL theory, for two simple examples, viz., (a) a one-dimensional scalar field theory, ie., the quantum dynamics of a massive particle, and (b) a scalar field in 2 spacetime dimensions.

%%%%%%%%%%%%%%%%%%%%%%%%%%%%%%%%%%%%%%%%%%%%%%%%%%%%%%
\subsection{One dimension (Particle dynamics)}
%%%%%%%%%%%%%%%%%%%%%%%%%%%%%%%%%%%%%%%%%%%%%%%%%%%%%%

In one spacetime dimension, i.e. quantum mechanics, the theory for a particle worldline $\gamma: \mb{R} \to {\cal M}$ is just a one-dimensional sigma model with target ${\cal M}$. We  consider a massive particle, with the action  (\ref{QMpart}). To do perturbation theory we expand the metric in inverse powers of the Planck mass $m_p$ as $g = \overline{g} + h/m_p$ about some background $\overline{g}$, so that
\be
\label{particleint}
S_o[g,q] \;\;=\;\;  S_0[\overline{g}, q] + \frac{1}{m_p} \int d^4x h_{\mu \nu} T^{\mu \nu} \;+\; O(h_{\mu\nu}^2)
\ee
where $S_0[\overline{g},q]$ is the action evaluated on the background metric, and the stress tensor is
\be
T_{\mu\nu}(x) = - m \int d\tau \frac{\dot{q}_{\mu} \dot{q}_{\nu}}{\sqrt{-\overline{g}_{\alpha\beta} \dot{q}^{\alpha} \dot{q}^{\beta}}} \delta(x - q(\tau))
\ee
which, if $\tau$ is proper time, becomes
\be
\label{particlestresstensor}
T_{\mu\nu}(x) = - m \int d\tau \dot{q}_{\mu} \dot{q}_{\nu} \delta(x - q(\tau)).
\ee

In what follows we will, for simplicity, assume that the backgound metric $\overline{g}^{\mu\nu}(x) = \eta^{\mu\nu}(x)$, ie., flat space. Then $S_0[\bar{g}, q] \rightarrow S_0[q]$, and the full gravitational action $S_G[g^{\mu\nu}]$ is replaced by the graviton action
\begin{equation}
S^{G}_o[h^{\mu\nu}] \;=\; {1 \over 2} \int h D_o^{-1} h
 \label{So-h}
\end{equation}
in which $D_o(q)$ is the flat space graviton propagator, given in the de Donder gauge by
\begin{equation}
D_o^{\mu\nu\alpha\beta}(q) \;=\; {1 \over q^2 + i\delta}(\eta^{\mu\alpha}\eta^{\nu\beta} + \eta^{\mu\beta}\eta^{\nu\alpha} - \eta^{\mu\nu}\eta^{\alpha\beta})
\end{equation}
where $\delta = 0^+$.

Let us write the CWL quantum-mechanical propagator as
\be
\mathcal{K}(x,x') = \sum_{n=1}^{\infty} {\cal K}_n(x,x')
\ee
where each term ${\cal K}_n(x,x')$ is
\begin{align}
{\cal K}_n(x,x')
\;\;&= \;\; \int Dh\; e^{ {i \over \hbar} S_o[h]}    \; \prod_k^n {1 \over n!} \int_{x'}^{x} Dq_k  \nonumber \\ & \times  \; \exp \left\{ {i \over n \hbar}  \sum_{i=1}^{n} (S_0[q_i] +   S_{int}[h,q_i] ) \right\}
\end{align}
to lowest non-vanishing corrections in the Planck mass. Here we dropped the overall normalization and defined $S_{int}$ from \eqref{particleint}.

%%%%%%%%%%%%%%%%%%%%%%%%%%%%%

\begin{figure}
\includegraphics[width=3.2in]{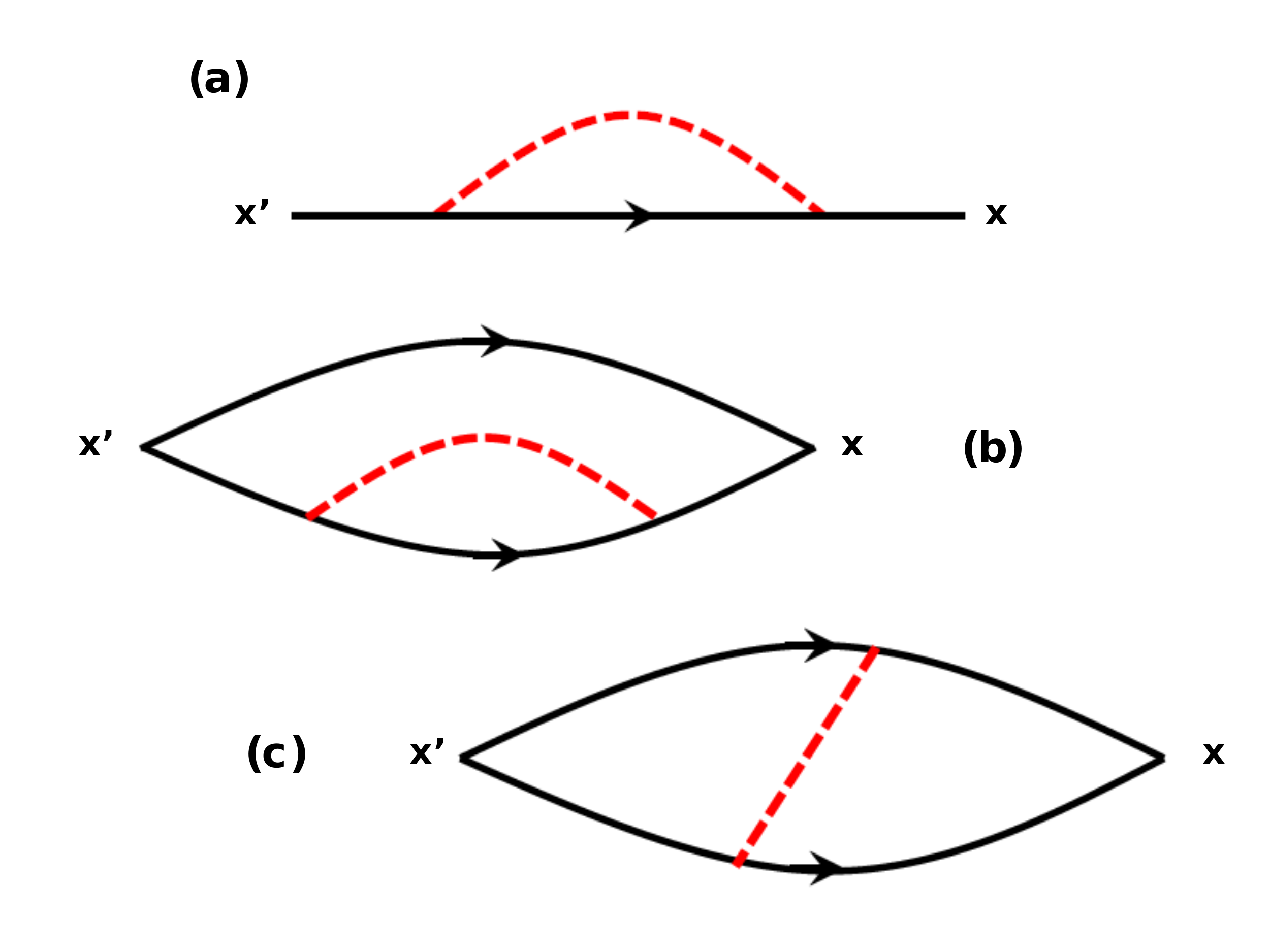}
\caption{\label{fig:Kn-fig} Contributions to the particle propagator $\mathcal{K}(x,x')$ for the summed version of CWL theory. In (a) we show the contribution $\Delta {\cal K}_1(x,x')$ in eqn. (\ref{DK1-xx}); in (b) the contribution $\Delta {\cal K}_2^A(x,x')$ of eqn. (\ref{K2A0}); and in (c) the contribution $\Delta {\cal K}_2^B(x,x')$ of eqn. (\ref{K2B0}). The particle paths are shown in black, the graviton paths in hatched red. }
\end{figure}

%%%%%%%%%%%%%%%%%%%%%%%%%%%%%

In what follows we will calculate these terms in ${\cal K}_n(x,x')$ up to $n=2$; the relevant graphs for this are shown in Fig. \ref{fig:Kn-fig}.

To lowest non-trivial order we have
\be
{\cal K}_1(x,y) = K_o(x,x') + \Delta {\cal K}_1(x,x')
\ee
where
\be
\label{particleG0}
K_o(x,x') = \int_{x'}^{x} Dq \;  \exp \left\{ {i \over \hbar} S_0[q] \right\}
\ee
is the ordinary propagator in the background metric $\overline{g}$, and $\Delta {\cal K}_1(x,x')$ is the conventional quantum gravity correction given by
\begin{eqnarray}
\Delta {\cal K}_1(x,x') &=&  - \frac{1}{m_p^2} \int {\cal D}h {\cal D}q \int d^4z d^4z' e^{ {i \over \hbar} \left(S_{o}^G[h] + S_o[q] \right) } \nonumber \\
& &\qquad  \times \; h_{\mu \nu}(z) T^{\mu \nu}(z) h_{\rho\lambda}(z') T^{\rho\lambda}(z')
 \label{DK1-xx}
\end{eqnarray}
ie., it is just the usual self-energy diagram for a particle coupled to gravitons.
There is no term linear in $1/m_p$; the Gaussian integral over ${\cal D}h$ vanishes if the number of gravitons is odd.

It is useful to evaluate $\Delta {\cal K}_1$ explicitly in background flat spacetime. Using \eqref{particlestresstensor}, we get
\begin{align}
\label{particleDG1}
\Delta {\cal K}_1(x,x') \;\;&=\;\; - \frac{i m^2}{4\pi^2 m_p^2} \int {\cal D}q d\tau d\tau' \nonumber \\
& \qquad\qquad \times  e^{ {i \over \hbar} S_0[q]} \; {\cal F}(q(\tau),q(\tau'))
\end{align}
where we define the function ${\cal F}(q(\tau),q(\tau'))$ of two arbitrary paths $q,q'$ at two arbitrary proper times $\tau,\tau'$ by
\begin{eqnarray}
{\cal F}(q(\tau),q'(\tau')) &=&  \mathbb{\cal P} \left[\frac{2 \left( \dot{q}(\tau) \cdot \dot{q}'(\tau') \right)^2 - \dot{q}^2(\tau) \dot{q}'^2(\tau')}{\left| q(\tau) - q'(\tau') \right|^2}\right] \nonumber \\
&& \qquad +\;\;  i\pi (\dot{q}^2)^2 \delta \left( \left| q(\tau) - q'(\tau') \right|^2 \right)\qquad
\end{eqnarray}
where $\mathbb{\cal P} [ \cdots ]$ indicates that the principal value integral is taken. The integrand in \eqref{particleDG1} is peaked around times $\tau,\tau'$ where $q(\tau) \approx q(\tau')$; when $\tau=\tau'$ the integral is ultraviolet divergent, as we see by writing $\tau' = \tau + \epsilon$ so that  $q^{\mu}(\tau') \approx q^{\mu}(\tau) + \epsilon \dot{q}^{\mu}(\tau)$; one then has
\begin{eqnarray}
\Delta {\cal K}_1(x,x') &=& - \frac{i m^2}{4\pi^2 m_p^2} \int {\cal D}q d\tau d\epsilon \; e^{ {i \over \hbar} S_o[q]} \nonumber \\ && \qquad \times \; \left[ \frac{\dot{q}^2(\tau)}{\epsilon^2} + \frac{i\pi (\dot{q}^2)^2}{\epsilon^2} \delta \left( \dot{q}^2(\tau) \right) \right] \qquad
\end{eqnarray}
plus finite terms.
%
%\be
%G(x,y) =
%\begin{fmffile}{particle-prop}
%\begin{fmfgraph*}(100,50)
%\fmfleft{i1,i2}
%\fmf{fermion}{i1,i2}
%\fmflabel{$x$}{i1}
%\fmflabel{$y$}{i2}
%\end{fmfgraph*}
%\end{fmffile}
%\ee

So far all these results (including the ultraviolet divergence in \eqref{particleDG1}) are just those in conventional quantum gravity. However at $n=2$
we get our first CWL-specific correction. We write the $n=2$ correction to the propagator as the sum of two terms, viz.,
\be
\Delta {\cal K}_2(x,x') = K_{2,0}(x,x') + \Delta {\cal K}_{2}(x,x')
 \label{DelK2}
\ee
in which the first term involves no gravitons, and takes the form
\begin{eqnarray}
K_{2,0}(x,x') &=&  \int_{x'}^x {\cal D}q {\cal D}q' \left[ 1 - \delta(q-q') \right]  \nonumber \\ && \qquad\qquad \times \; \exp {i \over 2 \hbar} ( S_o[q] + S_o[q'] )  \;\;\;\;
\end{eqnarray}
in which the functional $\delta$-function is just that term discussed above, ensuring that we don't double-count the lower-order contribution. Thus we have
\begin{eqnarray}
K_{2,0}(x,x') = \int_{x'}^x {\cal D}q {\cal D}q' \; e^{{i \over 2 \hbar} ( S_o[q] + S_o[q'] )} \;-\; K_o(x,x') \qquad
 \label{K20}
\end{eqnarray}
where the $K_o(x,x')$ term comes from the $\delta$-function contribution to the integrand.

The second term $\Delta {\cal K}_{2}(x,x')$  in (\ref{DelK2}) incorporates graviton interactions; written out in full it has the form
\begin{widetext}
\begin{align}
\Delta {\cal K}_2(x,x') & \;=\; - \frac{1}{m_p^2} \int {\cal D}h \; {\cal D}q {\cal D}q' \int d^4z d^4z' \; \left[ 1 - \delta(q-q') \right] \; e^{\;{i \over \hbar} \left[S_{o}^G[h] + \frac{1}{2} (S_0[q] + S_0[q'])\right]} \nonumber \\
& \qquad\qquad\qquad\qquad\qquad\qquad \times \; h_{\mu \nu}(z) \left( T_q^{\mu \nu}(z) + T_{q'}^{\mu\nu}(z) \right) h_{\rho\lambda}(w)  \left( T_q^{\rho \lambda}(w) + T_{q'}^{\rho\lambda}(w) \right).
\end{align}
This has four separate terms, which can be grouped into two pairs which are equal under a symmetry $q \leftrightarrow q'$ in the integrand. We write this as
$ \Delta {\cal K}_2 \;=\; 2 \Delta {\cal K}_2^A + 2 \Delta {\cal K}_2^B$, where the first term is
\begin{align}
\Delta {\cal K}_2^A(x,x') & \;=\; - \frac{1}{m_p^2} \int {\cal D}h \; {\cal D}q {\cal D}q' \int d^4z d^4z' \; \left[ 1 - \delta(q-q') \right] \; e^{\;{i \over \hbar} \left[S_{o}^G[h] + \frac{1}{2} (S_0[q] + S_0[q'])\right]} \nonumber \\
& \qquad\qquad\qquad\qquad\qquad\qquad \times \; h_{\mu \nu}(z) T_q^{\mu \nu}(z) h_{\rho\lambda}(w) T_q^{\rho \lambda}(w)
 \label{K2A0}
\end{align}
and we see that, just as we saw in the result for $K_{2,0}(x,x')$, the delta-function term keeps us from over-counting the lower-order terms:
\begin{align}
\Delta {\cal K}_2^A(x,x') \Big|_{\delta}  \;&=\;  \frac{1}{m_p^2} \int {\cal D}h {\cal D}q \int d^4z d^4z' \; e^{ {i \over \hbar} \left(S_{o}^G[h] + S_o[q] \right)} \; h_{\mu \nu}(z) \; T_q^{\mu \nu}(z) \; h_{\rho\lambda}(z') \; T_q^{\rho \lambda}(z') \nonumber \\ \;\;
 &\rightarrow \;\; -\Delta {\cal K}_1(x,x')
 \label{K2A1}
\end{align}
Likewise, the piece without the delta function again factors:
\begin{align}
\Delta {\cal K}_2^A(x,x') \Big|_{\text{no}\ \delta}  \;\;=\;\;\; - \frac{i m^2}{4 \pi^2 m_p^2} \int d\tau d\tau' \int_{x'}^x {\cal D}q' e^{ \frac{i}{2 \hbar} S_o[q']}
 \int_{x'}^x {\cal D}q \; e^{ \frac{i}{2 \hbar} S_o[q]} \; {\cal F}(q(\tau),q(\tau')).
 \label{K2A2}
\end{align}

The other term $\Delta {\cal K}_2^B$ describes processes that correlate the stress tensors on different paths $q$ and $q'$ via an internal graviton. Once again, the piece with the delta function simply gives $\Delta {\cal K}_2^B |_{\delta} = -\Delta {\cal K}_1$. The remainder is
\begin{align}
\Delta {\cal K}_2^B(x,x') \Big|_{\text{no}\ \delta}  \;\;=\;\;\; - \frac{i m^2}{4 \pi^2 m_p^2} \int d\tau d\tau' \int_{x'}^x {\cal D}q' e^{ \frac{i}{2 \hbar} S_o[q']}
 \int_{x'}^x {\cal D}q \; e^{ \frac{i}{2 \hbar} S_o[q]} \; {\cal F}(q(\tau),q'(\tau')).
  \label{K2B0}
\end{align}
\end{widetext}
This term describes a pair of worldlines interacting via gravity (note the difference from $\Delta G_2^A(x,x')$: the argument of ${\cal F}$ now involves $q'(\tau')$ instead of $q(\tau')$).
The kernel ${\cal F}$ is peaked for $\tau,\tau'$ such that $| q(\tau) - q'(\tau')| \approx 0$, ie., where the 2 paths approach each other very closely. Since we are including arbitrary pairs of paths $q,q'$ this can happen an arbitrary number of times.

All of the new physics appears first in this $n=2$ term. It was treated in detail for the non-relativistic limit (ie., where the particle velocity $\ll c$) in earlier papers \cite{stamp12,stamp15}. It is responsible for the ``path-bunching" effect described in these papers - it is this path-bunching that leads to the key observable differences from standard quantum mechanics, for massive particles.

%%%%%%%%%%%%%%%%%%%%%%%%%%%%%%%%%%%%%%%%%%%%%%%%%%%%%
\subsection{Two-dimensional scalar field}
%%%%%%%%%%%%%%%%%%%%%%%%%%%%%%%%%%%%%%%%%%%%%%%%%%%%%

The example of a scalar field on a 2-d spacetime is very useful because it allows us to consider, in a manageable way, the effects of both the spacetime topology and a non-trivial Faddeev-Popov determinant in a CWL theory. Let us recall some basic facts about this theory. The Einstein action for some spacetime manifold ${\cal M}$ of dimension $N$ is
\be
S_{G} = \int d^2x \sqrt{-g} R \;=\; \chi({\cal M}) \;\;=\;\; 2(1 - \text{genus}({\cal M}))
\ee
and is purely topological and independent of the metric $g$; for some fixed topology of ${\cal M}$, it is simply an overall rescaling in the generating functional, but in the full path integral we get a weighting over different topologies. The matter action
\be
S_M = \int_{\cal M} d^2x \sqrt{-g} g^{\mu \nu} G_{ab} \partial_{\mu} \phi^a \partial_{\nu} \phi^b.
\ee
is just the Polyakov action for bosonic string theory, ie., a linear $\sigma$-model with target space metric $G_{ab}$.

The 2-d Einstein equation of motion $T_{\mu \nu} = 0$,
giving us a relation between the matter field and the metric, and yielding classical constraints on the fields (eg., in string theory the string oscillations are constrained to be transverse to the string center of mass velocity). The theory possesses a lot of symmetry; we have (i) Global symmetries given by the isometry group of the target metric $G_{ab}$; (ii) worldsheet diffeomorphisms; and (iii) symmetry under Weyl/scaling transformations, ie., conformal transformations  $g_{\mu \nu}(x) \to \Omega^2(x) g_{\mu \nu}(x)$ of the metric, with $\phi$ transforming trivially (in conventional string theory the Weyl symmetry is anomalous, and cancelling this anomaly leads to the famous result $N=26$). Thus we have a gauge group $G = \text{Diff} \times \text{Weyl}$, which allows us to fix the metric $g$ to any given fiducial metric that we want; eg., $g^{\mu\nu} = \eta^{\mu\nu}$.

Consider now the CWL generating functional. Assuming (i) some fixed topology for $M$, and (ii) for simplicity that $G_{ab} = \delta_{ab}$, we have
\begin{eqnarray}
Q_n[J] &=&  \oint \frac{{\cal D}g}{V_{\rm Diff\times Weyl}} \prod_{a=1}^N D\phi^a_1 \cdots D\phi_n^a \nonumber \\
&& \qquad \times \; e^{ {i \over \hbar} \left( \sum_{i=1}^{n} S_M[g,\phi_i] + \int d^2x\  J_a(x) \phi^a_i \right)} \;\;\;\;
\label{CWL2dZ}
\end{eqnarray}
where the division by the volume of the full gauge group $V_{\rm Diff\times Weyl}$ should reproduce the Faddeev-Popov gauge fixing procedure with a relevant ghost determinant. This can be explicitly done in the same way as is usually done for a 2-d scalar theory of this kind; we let $g$ be some metric, let $\zeta = \zeta(x) = (v^{\mu}(x),\varphi(x))$ be the set of gauge parameters of the diffeomorphism and a Weyl transformation, and let $g^{\zeta}$ be the image of $g$ under this transformation. Then, fixing a fiducial metric $\hat{g}$, and ignoring Gribov ambiguities, for any metric $g$ there will be one $\zeta$ such that $g = \hat{g}^{\zeta}$. Now consider the integral over the gauge orbit of $\hat{g}$ given by
\be
1 = \Delta[g] \int D\zeta \delta(g - \hat{g}^{\zeta})
 \label{FPdef}
\ee
If we now insert this into \eqref{CWL2dZ} and take into account gauge invariance of the integration measures (over fields and group parameters), then because nothing in the integrand is $\zeta$-dependent, we observe cancellation of the group volume along with the removal of integration over the metric $g$ and get
\begin{eqnarray}
Q_n[J,\hat{g}] &=&  \oint \prod_{a=1}^N D\phi^a_1 \cdots D\phi_n^a \; \Delta[\hat{g}] \nonumber \\
&& \;\;\;\; \times \; e^{ {i \over \hbar} \left( \sum_{i=1}^{n} S_M[\hat g,\phi_i] + \int d^2x\  J_a(x) \phi^a_i \right)} \;\;\;\;\;\;
\label{CWL2dZ-fixed}
\end{eqnarray}
with the Faddeev-Popov determinant and matter action on the fiducial metric $\hat{g}$. To calculate $\Delta[\hat{g}]$ we write an infinitesimal gauge transformation in the vicinity of $\hat{g}$ as
\be
\hat{g}^\zeta_{\mu\nu} = \hat{g}_{\mu\nu}+2 \varphi \hat{g}_{\mu\nu} + \hat\nabla_{\mu} v_{\nu} + \hat\nabla_{\nu} v_{\mu}.
\ee
Writing the delta-function in (\ref{FPdef}) in terms of a functional Fourier transform one then has
\be
\Delta^{-1}[\hat{g}] = \int {\cal D}\varphi {\cal D}v {\cal D}\beta \; e^{  2 \pi {i \over \hbar} \int d^2x \sqrt{-\hat{g}} \beta^{\mu\nu}
(\hat{g}_{\mu\nu}-\hat{g}^\zeta_{\mu\nu})}\,.
\ee

The manouevres from here on are standard; after doing the Weyl integral $D\varphi$ (which enforces the tracelessness of the integration variable $\beta_{\mu\nu}$) we convert the c-number functions $v^{\mu}$ and $\beta_{\mu\nu}$ into a pair of anti-commuting Grassmann variables, viz., $v^{\mu} \to c^{\mu}$ and $\beta^{\mu\nu} \to b^{\mu\nu}$, and writing
\begin{align}
\begin{split}
\Delta[\hat{g}] & = \int Db Dc \; \exp \left\{ 4 \pi {i \over \hbar} \int d^2x \sqrt{-\hat{g}}  b^{\mu\nu} \hat\nabla_{\mu} c_{\nu} \right\} \\
& \equiv \int Db Dc \; \exp \left\{ {i \over \hbar} S_{\rm gh}[\hat{g},b,c] \right\}.
\end{split}
\end{align}
which defines the ghost action $S_{\rm gh}$, we finally end up with the $n$-th level contribution to the generating functional in the form
\be
Q_n[J,\hat{g}]  \;=\; \int Db Dc \prod_{a=1}^N D\phi^a_1 \cdots D\phi_n^a \; e^{{i \over \hbar} S_{\rm eff}}
\ee
with an effective action
\begin{eqnarray}
S_{\rm eff} &=& S_{\rm gh}[\hat{g},b,c] \nonumber \\
&& \;\;\; +
\sum_{i=1}^{n} S_M[\hat{g},\phi_i] + \int d^2x\  J_a(x) \phi^a_i
 \label{CWL2dZ-withghost}
\end{eqnarray}
which includes the ghost fields $c^{\mu}$ and $b^{\mu\nu}$.

At this stage, one can start computing correlation functions with the usual rules; the Feynman diagrams will simply include these ghost fields. One way to do this is to keep working with the string technology and do things in terms of a conformal field theory. Many interesting questions can then be addressed:
how one handles anomalous symmetries in a summed CWL framework (and whether or not the critical dimension $N=26$ of the bosonic string theory is changed), how to do operator product expansions; and so on.

However before doing any of this we need to address some more straightforward questions of internal consistency.

%%%%%%%%%%%%%%%%%%%%%%%%%%%%%%%%%%%%%%%%%%%%%%%%%%%%%%%%%%%%%
%%%%%%%%%%%%%%%%%%%%%%%%%%%%%%%%%%%%%%%%%%%%%%%%%%%%%%%%%%%%%

\section{Consistency tests for summed CWL}
 \label{sec:cons}

%%%%%%%%%%%%%%%%%%%%%%%%%%%%%%%%%%%%%%%%%%%%%%%%%%%%%%%%%%%%%
%%%%%%%%%%%%%%%%%%%%%%%%%%%%%%%%%%%%%%%%%%%%%%%%%%%%%%%%%%%%%

As already noted in the introduction, we need with any variant of CWL theory to make sure it passes all consistency tests - we need to check conservation laws in the dynamics, the classical limit, the non-interacting limit where the gravitational coupling $G \rightarrow 0$, and, if relevant, the renormalizability. In what follows we will find that the summation version of CWL theory passes most of these tests, but that there are problems in the classical limit.

%%%%%%%%%%%%%%%%%%%%%%%%%%%%%%%%%%%%%%%%%%%%%%%%%%
\subsection{Equations of motion}
 \label{sec:eom}
%%%%%%%%%%%%%%%%%%%%%%%%%%%%%%%%%%%%%%%%%%%%%%%%%%

A basic result in conventional QFT is that the classical equations of
motion hold as an operator equation. Thus suppose we consider an arbitrary
local QFT,
\be
\label{generalaction}
S_M = \int d^dx \sqrt{-g} \Li \left(\psi(x),\partial_{\mu}\psi(x),\ldots\right).
\ee
where $\psi(x)$ is some field (fermionic or bosonic), for which the
classical equation of motion is $\delta S/\delta \psi(x) = 0$. We can view
the classical equation of motion as an operator; define
\be
\Oi_{EOM}(x) = \delta S / \delta \psi(x)
\ee

so that the expectation value of this vanishes:
\begin{align}
\begin{split}
\braket{0 | \Oi_{EOM}(x) | 0} & = \int D\psi \ \frac{\delta S_M}{\delta
\psi(x)} \exp\left\{{i \over \hbar} S_M[\psi] \right\} \\
& = {i \over \hbar} \int D\psi \ \frac{\delta}{\delta \psi(x)} \exp\left\{{i \over \hbar} S_M[\phi]
\right\} \\
& = 0.
\end{split}
\end{align}
It is straightforward to show that one can further insert any number of local operators in the correlation function and still obtain a vanishing result.

In the summed version of CWL theory, the matter equations of motion are likewise satisfied as operator equations. However, there is a fundamental difficulty with the gravitational equations of motion. This provides the first hint of the serious difficulty with the semiclassical $\hbar \to 0$ limit, to be discussed in the next section.

Let us demonstrate these statements in detail. Fix a CWL level $n$, and consider performing the same basic technique of differentiating with respect to $\delta/\delta \psi$. We have
\begin{widetext}
\begin{align}
\begin{split}
0 & \;=\;  \int Dg D\psi_1 \cdots D\psi_n \sum_{i=1}^{n} \frac{\delta}{\delta
\psi_i(x)}  \exp\left\{{i \over \hbar} (S_g[g] + \frac{i}{n \hbar} \sum_{i=1}^{n} S_M[g,\psi_i]
)\right\}  \\
& \;=\; \int Dg D\psi_1 \cdots D\psi_n \; \frac{i}{n \hbar} \sum_{i=1}^{n} \frac{\delta
S_M}{\delta \psi_i(x)}   \exp\left\{{i \over \hbar} (S_g[g] + \frac{i}{n \hbar} \sum_{i=1}^{n}
S_M[g,\psi_i] )\right\}  \\
& \;=\; \braket{0 | \Oi_{EOM}(x) | 0}_n.
\end{split}
\end{align}
Here the second line is defined by eqns. (\ref{Qn-gJa}) and (\ref{generalopcorr}), (\ref{FxQJ}) above, with again
$\Oi_{EOM}(x) = \delta S/\delta \psi(x)$. In other words, the equations of motion hold as an operator separately at each CWL level $n$. Again, one can easily generalize this to include an arbitrary set of other local operator insertions. Thus, so far so good.

However, consider now the Einstein equations. These would naively be obtained via a similar manouevre, taking the derivative with respect to $\delta/\delta g$ under the integral. This yields
\begin{align}
\begin{split}
0 & \;=\;  \int Dg D\psi_1 \cdots D\psi_n \; \frac{\delta}{\delta g_{\mu\nu}(x)}
  \exp\left\{{i \over \hbar} (S_g[g] + \frac{i}{n \hbar} \sum_{i=1}^{n} S_M[g,\psi_i] )\right\}
 \\
 & \;=\;  \int Dg D\psi_1 \cdots D\psi_n \; {i \over \hbar} \left( \frac{\delta S_g}{\delta
g_{\mu\nu(x)}} + \frac{1}{n} \sum_{i=1}^{n} \frac{\delta
S_M[g,\psi_i]}{\delta g_{\mu\nu}(x)} \right) \exp\left\{{i \over \hbar} (S_g[g] + {i \over n \hbar}
\sum_{i=1}^{n} S_M[g,\psi_i] )\right\}.
\end{split}
\end{align}
\end{widetext}

Comparing again to eqtns. (\ref{Qn-gJa}), (\ref{generalopcorr}), and (\ref{FxQJ}) above, we see that this is of the
form $\braket{ \Oi_{GR}(x) }$, but with $\Oi_{GR} = n \delta S_g/\delta g +
\delta S/\delta g$; thus, it is off by a relative factor of $n$. In other words, to put this as the correct operator equation, one would need to rescale the metric by multiplying by $n$. But then this would cause the quantum fluctuations of the metric to dominate the path integral at high $n$. This is our first hint that the semiclassical expansion is peculiar; we will examine this issue in more detail below.

%%%%%%%%%%%%%%%%%%%%%%%%%%%%%%%%%%%%%
\subsection{Noether's theorem}
%%%%%%%%%%%%%%%%%%%%%%%%%%%%%%%%%%%%%

In standard QFT one considers transformations that preserve the product of the measure and weight in the path integral, ie., local transformations $\phi \to \phi'$ of all the fields such that
\be
\label{generalsymmetry}
D\phi e^{{i \over \hbar} S[\phi]} = D\phi' e^{{i \over \hbar} S[\phi']}.
\ee
It is not immediately obvious how to think of this in CWL, because it would need to hold term by term in $n$, which is certainly not implied by \eqref{generalsymmetry}. We will assume that the measure $D\phi = Dg D\psi_1 \cdots D\psi_n$ is \emph{invariant} all by itself. Clearly, if these conditions hold, then any correlation function of the form \eqref{generalopcorr} is invariant.

The statement of Noether's theorem is then the same in CWL as in ordinary QFT. It is that there is a local current $j^{\mu}(x) = j^{\mu}(g(x),\partial g(x), \ldots, \psi(x),\partial \psi(x),\ldots)$ built out of the metric, matter fields, and their derivatives, which is covariantly conserved, ie., that
\be
\braket{ 0 | \nabla_{\mu} j^{\mu}(x) \Oi_1(x_1) \cdots \Oi_N(x_N) | 0 } = 0
\ee
for any collection of local functions $\Oi_a(x_a)$, as long as none of the $x_a$ coincide with $x$.

To show this we follow fairly standard manouevres: consider the transformation
\begin{align}
\begin{split}
\label{non-symmetry}
\psi(x) & \to \psi'(x) = \psi(x) + \epsilon \rho(x) \delta \psi(x) \\
g_{\mu\nu}(x) & \to g_{\mu\nu}'(x) = g_{\mu\nu}(x) + \epsilon \rho(x) \delta g_{\mu\nu}(x).
\end{split}
\end{align}
This is not a symmetry of the action, but if $\rho(x)$ were a constant then it would be - these transformations would then be exact global invariances of the Lagrangian. Thus the variation of the action under this transformation must be proportional to a total derivative of $\rho$, ie.,
\begin{align}
\label{non-symmetry-variation}
\begin{split}
&S[g',\{ \psi'_k \}] =  S_{G}[g'] + \frac{1}{n} \sum_{k=1}^n S_M[g',\psi'_k] \\
& \;\; = S[g, \{ \psi_k \}]  + \frac{i \epsilon}{2\pi n} \sum_{i=k}^{n} \int d^dx \sqrt{-g}j^{\mu}(g,\psi_k) \partial_{\mu} \rho.
\end{split}
\end{align}
where as usual the variations $\delta\psi, \delta g$ are arbitrary functions which vanish sufficiently fast at infinity so that we can neglect these boundary terms, and $\epsilon \ll 1$ is a small constant parameter. The $i/2\pi$ factor is for later convenience and can be absorbed into the definition of $j^{\mu}$.

Note that the explicit formula for $j^{\mu}(g,\psi)$ follows from just the action with a single copy of the matter. Indeed, for a given copy, the variation under \eqref{non-symmetry} is
\be
\frac{1}{n} \delta_{\epsilon}S[g, \{ \psi_k \}]  = \frac{i \epsilon}{2 \pi n} \int d^dx \sqrt{-g} j^{\mu}(g,\psi) \partial_{\mu} \rho
 \label{deleS}
\ee
and we can re-write the exponent in the path integral as a sum of $n$ such terms.

Now, let us consider the general expectation value \eqref{generalopcorr} for a product of local functions of the field $\psi(x)$. If we perform the transformation \eqref{non-symmetry} and take $\rho(x)$ to vanish everywhere except some small region $R$ around $x$, with all the other insertions $x_a \notin R$, then this is a symmetry of \eqref{generalopcorr}. This means that we can compute $F$ either using $g,\psi$ or $g',\psi'$. In particular, we can do so term-by-term at each $n$. Then, making a Taylor expansion of the exponential in the path integral, and integrating the $\partial_{\mu} \rho$ term by parts, we have
\begin{widetext}
\begin{align}
\label{noetherderiv}
\begin{split}
0 & \;=\; F'_n - F_n \\
& \;=\; \frac{\epsilon}{2\pi n} \int Dg \; D\psi_1 \cdots D\psi_n \Delta[g] \int d^dx  \sqrt{-g} \rho(x) \sum_{i=1}^n\nabla_{\mu} j^{\mu}(g(x),\psi_i(x)) \;  \sum_{i=k}^{n} \prod_{a=1}^{N} \Oi_a(\psi_k(x_a)) \; e^{ {i \over \hbar} S[g, \{ \psi_k \}]}  \\
& \;=\; \int d^dx \sqrt{-g} \rho(x) F^{j}_n(x,x_1,\ldots,x_N).
\end{split}
\end{align}
where $F^{j}_n = F^{j}_n(x,x_1,\ldots,x_N)$ is the $n$th term in the CWL expression for the correlator \eqref{generalopcorr} with an insertion of $\nabla_{\mu} j^{\mu}(x)$, viz.
\begin{align}
\begin{split}
& F^j_n(x,x_1,\ldots,x_N)\; \;= \int Dg \; D\psi_1 \cdots D\psi_n \Delta[g] \; \sum_{i=1}^{n}  \nabla_{\mu} j^{\mu}(g(x),\psi_i(x))  \sum_{i=1}^{n} \prod_{a=1}^N \Oi_a(\psi_i(x_a))  \; e^{ {i \over \hbar} S[g, \{ \psi_k \}]}
\end{split}
\end{align}
\end{widetext}
Since $\rho(x)$ is arbitrary other than its restriction to the region $R$, one concludes from \eqref{noetherderiv} that $F^{j} = \sum_{n=1}^{\infty} F_n^j = 0$. Thus we have
\be
\braket{ 0 | \nabla_{\mu} j^{\mu}(x) \Oi_1(x_1) \cdots \Oi_N(x_N) | 0 } = 0,
\ee
which verifies Noether's theorem for CWL theory.

%%%%%%%%%%%%%%%%%%%%%%%%%%%%%%%%%%%%%%%%%%%%%%%%%%%%%%%%%%%%%
\subsection{Classical Limit}
 \label{sec:classLim}
%%%%%%%%%%%%%%%%%%%%%%%%%%%%%%%%%%%%%%%%%%%%%%%%%%%%%%%%%%%%%

We have already looked above at the structure of low-order perturbation theory, ie., at the limit where the corrections to the behaviour of the system in the absence of gravity are small. Notice that this limit not only switches off the CWL correlations, but it also switches off the conventional corrections to non-gravitational behaviour, ie., it suppresses completely the coupling to gravitons. In this sense the behaviour of perturbation theory does not allow us to separate CWL effects from ordinary quantum gravitational effects.

Another important limit is the classical limit, when $\hbar \rightarrow 0$. In this paper we will not try to develop any kind of systematic semiclassical expansion in powers of $\hbar$, for either summed or product CWL theories. However we can look to see how sensible the classical limit is for the 2 theories.

There is a problem with the classical limit of the summed version of CWL theory, in the form given in eqtn. (\ref{Q-sch1}); it comes from the form assumed for $Q_n[g,J]$. To see this, consider a more general form for $Q_n[g,J]$, given for a scalar field by
\begin{equation}
Q_n[J,g] = \prod_{k=1}^n \oint {\cal D}\phi_k \; e^{{i \over v_n \hbar } \sum_k \;   (S_M[\phi_k, g^{\mu \nu}] + \int J \phi_k)}
 \label{Qn-gJ2}
\end{equation}
instead of that given in (\ref{Qn-gJ}). Here, instead of the factor $1/n \hbar$ which appears in front of the exponent in the matter action in (\ref{Qn-gJ2}), we have a factor $1/v_n \hbar$, where $v_n$ is an arbitrary function of the integer $n$. In the case chosen for all our perturbative calculations, viz., $v_n = n$, we notice that there are cancellations involved which prevent any kind of double-counting of terms (cf. the remarks after eqtns. (\ref{K20}), (\ref{K2A0}), and (\ref{K2A2}) in the previous section). Any change from the specification $v_n = n$ ruins all these cancellations, and invalidates the perturbative expansion.

However the assumption that $v_n = n$ means that the classical limit is rather peculiar. When $\hbar = 0$, we do have the correct classical theory; but when $\hbar \neq 0$, we see that at order $n$, the effect of quantum fluctuations is multiplied by $n$ as compared to standard QFT. Effectively, $\hbar$ is multiplied by a factor $n$ at $n$-th order. Physically, this means that the effect of quantum fluctuations becomes very large for $n$-tuples of paths correlated by gravity, and indeed divergent in the $n \rightarrow \infty$ limit.

While such a behaviour does not necessarily render the theory invalid (to show this we would need to know how to sum over all $Q_n$), it is certainly a motivation for seeking an alternative formulation of the CWL idea, to which we now turn.

%%%%%%%%%%%%%%%%%%%%%%%%%%%%%%%%%%%%%%%%%%%%%%%%%%%%%%%%%%%%%
%%%%%%%%%%%%%%%%%%%%%%%%%%%%%%%%%%%%%%%%%%%%%%%%%%%%%%%%%%%%%

\section{CWL Theory: Product Version}
 \label{sec:CWLprod}

%%%%%%%%%%%%%%%%%%%%%%%%%%%%%%%%%%%%%%%%%%%%%%%%%%%%%%%%%%%%%
%%%%%%%%%%%%%%%%%%%%%%%%%%%%%%%%%%%%%%%%%%%%%%%%%%%%%%%%%%%%%

So far we have worked exclusively with the summed version of CWL theory. However, as noted in the introduction, one can develop an alternative version of CWL theory, in which one sums not over sets of correlated paths in the generating functional, but rather its logarithm. In this section we introduce arguments which lead us to this kind of ``product CWL" theory, and also show that, unlike the summed version, it clearly has a sensible smooth classical limit.

%%%%%%%%%%%%%%%%%%%%%%%%%%%%%%%%%%%%%%%%%%%%%%%%%%%%%%%%%%%%%
\subsection{Uncorrelated Worldlines and Matter field Replicas}
 \label{sec:Z-UWL}
%%%%%%%%%%%%%%%%%%%%%%%%%%%%%%%%%%%%%%%%%%%%%%%%%%%%%%%%%%%%%

The easiest way to see how to build up a product version of the CWL theory is to begin by imagining a theory of $n$ uncoupled multiple copies of a quantum matter field. We can consider each of these to exist in a separate ``replica" of the universe; any attempt to switch on gravitational correlations between them then produces a CWL theory. However as we shall see there are different ways of doing this.

We begin by ignoring gravity completely, and consider a scalar field $\phi(x)$ which exists in multiple copies $\phi_k(x)$, with $k = 1, \cdots, n$, giving a theory defined by a generating functional
    \begin{eqnarray}
    {\cal Z}^U[\,J\,]  &=&  \prod\limits_{n=1}^\infty \prod\limits_{k=1}^n\int D\phi_k\,e^{{i \over \hbar}(S_m[\,g,\phi_k^{(n)}\,] \;+\; \int \phi_k\frac{J}{c_n})} \nonumber \\    &=&
    \prod\limits_{n=1}^\infty \,
    \left(Z\Big[\,\frac{J}{c_n}\,\Big]\right)^n,
    \end{eqnarray}
where the generator $Z[\,J\,]$ for a single copy of the field is
    \begin{eqnarray}
    &&Z[\,J\,]=\int D\phi\,e^{{i \over \hbar}(S_m[\,g,\phi\,]\;+\;\int J\phi)}
    \end{eqnarray}
and we have introduced a function $c_n$ which will act as a regulator of divergences generated by this infinite product over $n$, which will require that $c_n$ grow sufficiently rapidly with $n$ (note that although $c_n$ is analogous to the factor $v_n$ introduced in the last section, we shall see it plays a rather different role here). The matter action is a functional of both the matter field $\phi(x)$ and a background metric field $g^{\mu \nu}(x)$, which for the moment we will assume to be frozen in some specific configuration.

We see that $Z[\,J\,]$ is a conventional generating functional for a single quantum scalar field, possessing the connected correlators
    \begin{eqnarray}
    &&\langle\,\phi\,\rangle=
    \frac\delta{\delta J}\ln Z[\,J\,]\Big|_{\,J=0} \nonumber \\
    &&\langle\,\phi(x_1)...\phi(x_l)\,
    \rangle_{\rm c}=\left.\frac{\delta^l\ln Z[\,J\,]}{\delta J(x_1)...\delta J(x_l)}\,\right|_{\,J=0} \;\;\;\;\;\;
    \label{Z-corr}
    \end{eqnarray}
in the usual way.

The generating functional ${\cal Z}^U[\,J\,]$ for the multi-field uncorrelated replicas is then an infinite product over $n$ replicas of $Z$, with $n \rightarrow \infty$. Since we have
    \begin{eqnarray}
    &&\frac\delta{\delta J}\ln{\cal Z}^{U}[\,J\,]=\sum\limits_{n=1}^\infty\,
    \frac{n}{c_n}\frac\delta{\delta j}
    \ln Z\big[\,j\,\big]_{\,j\,= {J \over c_n}},  \;\;\;\;\;
    \end{eqnarray}
then the correlation functions for the infinite product must then be defined as
    \be
    \frac\hbar{i}\frac\delta{\delta J}\ln{\cal Z}^{U}[\,J\,]\Big|_{\,J=0}=
    \langle\,\phi\,\rangle\,\sum\limits_{n=1}^\infty\,
    \frac{n}{c_n}, \nonumber
    \ee
    \be
    \left.\left(\frac\hbar{i}\right)^l\frac{\delta^l\ln {\cal Z}^{U}[\,J\,]}{\delta J(x_1)...\delta J(x_l)}\,\right|_{\,J=0}=
    \langle\,\phi(x_1)...\phi(x_l)\,
    \rangle_{\rm c}\; \sum\limits_{n=1}^\infty\,
    \frac{n}{c^l_n} \;\;\;\qquad
    \label{ZU-corr}
    \ee
where $\langle\,\phi\,\rangle$ and $\langle\,\phi(x_1)...\phi(x_l)\,
\rangle_{\rm c}$ are the usual QFT mean field and connected correlators respectively, as defined in (\ref{Z-corr}). If $c_n$ grows with $n$ fast enough the sums in (\ref{ZU-corr}) are convergent, and these relations can be interpreted as definitions of quantum correlators. For uncorrelated paths they then generate the usual quantum averages. Thus, as we might expect, in this uncorrelated ``replica" theory we just get back standard QFT.

%%%%%%%%%%%%%%%%%%%%%%%%%%%%%%%%%%%%%%%%%%%%%%%%%%%%%%%%%%%%%
\subsection{Switching on Gravitational Correlations}
 \label{sec:G-corr}
%%%%%%%%%%%%%%%%%%%%%%%%%%%%%%%%%%%%%%%%%%%%%%%%%%%%%%%%%%%%%

In any kind of CWL theory, a key part of the theory is the idea that one generates correlations between worldlines by switching on the gravitational interactions, which then implies an additional integration over the metric field - in the language used above, we correlate the replica fields. There are a number of ways that one might do this within the product structure we have adopted - here we will look at two ways, one of which uses a single metric field, the other of which introduces replica metric fields as well. We will see that the latter method is preferable.

\vspace{3mm}

{\bf (i) Single Metric Field:} The simplest way of introducing gravitational correlations between the separate matter field replicas is to introduce a single dynamic metric field coupled to each of these replicas, ie., we unfreeze the background field $g^{\mu\nu}(x)$ in the uncorrelated theory, and give it dynamics.

Schematically we then have
    \begin{eqnarray}
    {\cal Z}^{U}[\,g,J\,] &\to & \; {\cal Z}^{CWL}[\,J\,] \nonumber \\
     &\sim&
    \int Dg \; \Delta(g) \; e^{{i \over \hbar}S_G[\,g\,]}\,{\cal Z}^{U}[\,g,J\,]  \qquad  \label{grav}
    \end{eqnarray}
where $S_G[\,g\,]$ is the gravitational action, and $\Delta(g) = \det|\Xi(g)|$ is the determinant of the Faddeev-Popov gauge operator $\;\hat{\Xi}(g)$, the inverse propagator of ghost fields. Written out in full, we then have
\begin{widetext}
    \begin{align}
    {\cal Z}^{CWL}[\,J\,] \;\;=\;\; \int Dg \; \Delta(g) \,\left(\,\prod\limits_{n=1}^\infty D\phi_1^{(n)}\, D\phi_2^{(n)}...D\phi_n^{(n)}\right)
    \exp\left( {i \over \hbar} \Big(\,S_G[\,g\,]
    +\sum\limits_{n=1}^\infty\sum\limits_{i=1}^n
     S_m[\,g,\phi_i^{(n)}\,]
    \;+\;\int \frac{J}{c_n}\phi_i^{(n)}\Big)\right).
     \label{Z0}
    \end{align}
\end{widetext}
This multiple (infinite) field path integral implies a conventional QFT with the infinite set of quantum fields
    \begin{align}
    \varPhi^A=\phi_k^{(n)}(x); \quad k=1,2,...n,\quad n=1,2,...\infty.
    \end{align}
as well as the gravitational field $g^{\mu\nu}(x)$. This theory can therefore be renormalized by a set of local counterterms, and it has standard renormalizability or non-renormalizability properties, depending on the form of the matter and gravitational actions $S_m[\,g,\phi\,]$ and $S_G[\,g\,]$.

Given a CWL generating functional of this form, we will then have the CWL correlators {\em defined} by the relations
    \begin{eqnarray}
    &&\!\!\!\!\langle\,\phi(x)\,\rangle^{CWL}=
    \left(\sum\limits_{n=1}^\infty\,
    \frac{n}{c_n}\right)^{-1}\!\frac\hbar{i}
    \frac{\delta\ln{\cal Z}^{CWL}[J]}{\delta J(x)}\Big|_{J=0},\\
    &&\!\!\!\!\langle\,\phi(x_1)...\phi(x_l)\,
    \rangle^{CWL}_{\rm c}=\left(\sum\limits_{n=1}^\infty\,
    \frac{n}{c^l_n}\right)^{-1}   \nonumber \\
    && \qquad\qquad\qquad\times \left.
    \left(\frac\hbar{i}\right)^l\frac{\delta^l
    \ln {\cal Z}^{CWL}[\,J\,]}{\delta J(x_1)...\delta J(x_l)}\,\right|_{\,J=0} \;\;\qquad                         \label{CWLcorrelator}
    \end{eqnarray}
With gravity switched off, ${\cal Z}^{CWL}$ goes over into ${\cal Z}^{U}$, and these averages reduce to conventional QFT ones. Thus the ``correspondence principle", whereby as the gravitational coupling $G \rightarrow 0$ the limit of this theory becomes the flat space QFT, is trivially enforced.

It is important to note here that since the source $J(x)$ is coupled not to the individual quantum fields $\phi^A$, but to their particular (and infinite series) sum, ie., to
    \begin{align}
    {\cal O}=\sum\limits_{n=1}^\infty\frac1{c_n}\sum\limits_{i=k}^n \phi_k^{(n)}
    \end{align}
the S-matrix produced by this generating functional is not unitary. It is unitary in the Hilbert space of all $\phi^A$, but not on the subspace associated with these particular collective variables $\cal O$.

So far what we have sketched here seems like a perfectly consistent CWL theory. There is however a difficulty with this construction. If we attempt to build a semiclassical expansion for (\ref{Z0}), we get an infinite set of Einstein and matter field equations for the saddle-point configuration, given by
    \begin{align}
    &\frac{\delta S_G[\,g\,]}{\delta g}
    +\sum\limits_{n=1}^\infty\sum\limits_{k=1}^n
    \frac{\delta S_m[\,g,\phi_k^{(n)}]}{\delta g}=0,\\
    & \frac{\delta S_m[\,g,\phi_k^{(n)}]}{\delta\phi_k^{(n)}}
    -\frac{J}{c_n}=0.
    \end{align}
We see that the infinite sum over the matter stress tensors $T^{\mu\nu}[\,g,\phi_i^{(n)}]=2\delta S_m[\,g,\phi_i^{(n)}]/\delta g_{\mu\nu}$, for each individual field $\phi_k^{(n)}$, then gives an unregulated divergence in the Einstein equation for $g$. At first glance one might think of suppressing this by inserting an $n$-dependent coefficient $\tilde{c}^{-1}_n$ analogous to $c^{-1}_n$, this time multiplying $S_m[\,g,\phi_i^{(n)}]$, with $\tilde{c}_n)$ also growing sufficiently rapidly with $n$ to remove the divergence. However this does not work, because the replacement $S_m[\,g,\phi_i^{(n)}]\to S_m[\,g,\phi_i^{(n)}]/\tilde{c}_n$ amounts to introducing an effective $\hbar \propto \tilde{c}_n$, and so would then result in divergent radiative/quantum fluctuation corrections as $\tilde{c}_n\to\infty$.

\vspace{3mm}

{\bf (ii) Replication of Gravitational Fields:} To deal with the above problem, we introduce a set of metric fields $g_n$, with one such field for each tower of matter fields $\phi_k^{(n)}$, where again $i=1,2,...n$. We also rescale the $n$-th gravitational action by a factor $n$, for reasons discussed below, to get
    \begin{eqnarray}
    &&\mathbb{Q}[J] \;=\;\prod\limits_{n=1}^\infty \int Dg_n \;\Delta(g_n) \; e^{{in \over \hbar} S_G[\,g_n\,]} \nonumber \\ &&\qquad\qquad\qquad \times \prod\limits_{i=1}^n\int D\phi_i^{(n)}\,e^{{i \over \hbar} (S_m[\,g_n,\phi_i^{(n)}\,]
    \;+\;\int \frac{J}{c_n}\,\phi_i^{(n)})}\nonumber\\
    &&\qquad=
    \prod\limits_{n=1}^\infty
    \int Dg_n \;\Delta(g_n) \; e^{{in \over \hbar} S_G[\,g_n\,]}\,
    \left(Z\Big[\,g_n,
    \frac{J}{c_n}\,\Big]\right)^n  \qquad   \label{bbQ-J}
    \end{eqnarray}
This preserves all the advantages of the product version of CWL over the summed version, but drastically changes the saddle-point equations from those obtaining when we only have a single metric field; these equations now read
    \begin{eqnarray}
    &&n\,\frac{\delta S_G[\,g_n\,]}{\delta g_n}
    +\sum\limits_{i=1}^n
    \frac{\delta S_m[\,g_n,\phi_i^{(n)}\,]}{\delta g_n}=0,\\
    &&\frac{\delta S_m[\,g_n,\phi_i^{(n)}\,]}{\delta\phi_i^{(n)}}-{J \over c_n}=0
    \end{eqnarray}
Under the same boundary conditions for all $\phi_k^{(n)}$, the solutions for these equations coincide for all $k$ at the level of the $n$-th tower, ie., $\phi_k^{(n)}=\phi^{(n)}$. This means that all the $n$ matter stress tensors in the Einstein equations coincide too, and the overall coefficient $n$ cancels out to give
    \begin{eqnarray}
    \frac{\delta S_G[\,g_n\,]}{\delta g_n}
    +\frac{\delta S_m[\,g_n,\phi^{(n)}\,]}{\delta g_n}=0,
    \end{eqnarray}
so that at the level of any tower $n$ we get the usual Einstein equation, now sourced by the stress tensor of a {\em single} matter field $\phi^{(n)}$. Moreover, in the absence of sources, ie., with $J=0$, the background fields $g_n$ and $\phi^{(n)}$ satisfy the same set of equations and, therefore, coincide for all different $n$, ie., $\phi^{(n)}=\phi_0$, and $g_n=g_0$. The saddle points are then the same for every tower, which is very useful.

An important distinction between (\ref{bbQ-J}) and (\ref{Z0}) is that $\mathbb{Q}[J]$ incorporates gravitational correlations (in connected graphs) within each level $n$, but {\it not} between different $n$'s. This is because (\ref{bbQ-J}) factorizes, ie.,
    \begin{eqnarray}
    &&\mathbb{Q}[J]=\prod\limits_{n=1}^\infty {\cal Q}_n[\,J\,] \nonumber \\
    &&{\cal Q}_n[\,J\,]=
    \int Dg\,\Delta(g) \, e^{{in \over \hbar} S_G[\,g\,]}\,\left(Z\Big[\,g,
    \frac{J}{c_n}\,\Big]\right)^n   \qquad  \label{bbQ-J1}
    \end{eqnarray}
and its logarithm -- the generator of connected graphs -- is just a sum of contributions of single $g$ integrals. Thus we do not have correlations between $g_n$ and $g_m$ unless $n=m$. 

This last property is important, because all theories of two and more coupled metrics were for a long time believed to be non-unitary \cite{Cutler-Wald,Boulware_Deser,Damour_et_al}. Only recently has a special generally covariant bi-metric model been found that circumvents the unitarity problem (including the issue of nonlinear Boulware-Deser ghost \cite{Boulware_Deser}), for only two nonlinearly interacting metrics \cite{dRGT,Hassan-Rosen}. As we need more than two metric fields for CWL, and do not want to have any extra source of non-unitarity in the theory beyond the one established via CWL, the lack of correlation between different metric copies is a key feature of the product CWL structure.

This is important, because all known theories of two and more coupled metrics are non-unitary \cite{polyakov}, and we do not want to have any extra source of non-unitarity in the theory beyond the one established via CWL.

The rationale for the introduction of the factor $n$ multiplying $S_G[\,g_n\,]$ is less obvious. One could also imagine, eg., dividing $S_m[\,g_n,\phi^{(n)}\,]$ by $n$. The reason for the multiplicative factor is that it leaves the quantum fluctuation effects on the matter fields unenhanced at high $n$, and at the same time corresponds to a rescaling of the gravitational coupling constant for the $n$-th metric $g_n$, such that $G\to G/n$. This then weakens the gravitational coupling for growing $n$ and thereby reduces quantum fluctuation effects in $g_n$ (since $\hbar$ and $m_p^{-2}$ are transformed to effective couplings $\hbar/n$ and $1/nm_p^{2}$ respectively, at the level of the $n$-th tower). As we shall see in a more detailed analysis of product CWL theory, this also has the advantage of giving us a well-controlled semiclassical expansion around the classical limit, applicable for all $n$.

Eqn. (\ref{bbQ-J1}) summarizes the basic structure of what we are calling ``product CWL" theory. The obvious next step would be to work out the entire consequences of this form for $\mathbb{Q}[J]$. This is of course a fairly lengthy task, since we have to start from the beginning - showing how to do perturbative expansions for correlation functions and propagators, how to do semiclassical expansions and renormalization, and understand how gauge invariance and Ward identities are satisfied. Beyond this one would like to understand general questions about renormalizability, background field methods, anomalies, and infrared properties, as well as more concrete questions about, eg., how to do scattering calculations.  This is a big job, and will be done in subsequent papers.

However we can exhibit here the structure of perturbation theory for product CWL theory, and compare it with that found for summer CWL theory; this turns out to be very illuminating, and we turn to it now.

%%%%%%%%%%%%%%%%%%%%%%%%%%%%%%%%%%%%%%%%%%%%%%%%%%%%%%%%%%%%%%%%
%%%%%%%%%%%%%%%%%%%%%%%%%%%%%%%%%%%%%%%%%%%%%%%%%%%%%%%%%%%%%%%%%%%%%%%%%%%%

\section{Perturbation expansions for Product version of CWL theory}
\label{sec:pertProd}

%%%%%%%%%%%%%%%%%%%%%%%%%%%%%%%%%%%%%%%%%%%%%%%%%%%%%%%%%%%%%%%%
%%%%%%%%%%%%%%%%%%%%%%%%%%%%%%%%%%%%%%%%%%%%%%%%%%%%%%%%%%%%%%%%%%%%%%%%%%%%

It is straightforward to compare the structure of perturbative expansions in product CWL theory with that found above in summed CWL theory. The perturbative expansions will demonstrate an essential feature of product CWL theory, viz., how the inter-path correlations manifest themselves, and cause a breakdown of the superposition principle. That this should happen is not obvious from the product form in eqn. (\ref{bbQ-J1}), but a calculation of low order diagrams in the theory makes it clear.

In this section we will again compute to lowest non-trivial order in perturbation theory in gravitons, i.e. to order $1/m_p^2$. In treating the summed CWL theory we calculated this expansion for particle and 2-d scalar field propagators; here, to be specific we again pick the example of a scalar field, and find the 2-point and 4-point correlation functions for this field.  We will see that any process in product CWL can be calculated in terms of conventional Feynman diagrams, but with rather different combinatoric factors than a calculation in conventional quantum gravity.

The goal, as in the calculations of section \ref{sec:pertProd}, is to display the basic structure via simple ``brute force" computations. Once this is done it is clearly preferable to have a formal loop expansion at our disposal - in future papers we do this, and give the 1-loop renormalization of the field strength and mass in product CWL theory.

%%%%%%%%%%%%%%%%%%%%%%%%%%%%%%%%%%%%%%%%%%%%%%%%%%%%%%%%%
\subsection{Perturbation expansions for Correlators}
 \label{sec:prodCWL-corr}
%%%%%%%%%%%%%%%%%%%%%%%%%%%%%%%%%%%%%%%%%%%%%%%%%%%%%%%%%

As discussed already in the introduction, the field correlators
in any CWL theory are given directly from the generating
functional. Since we are dealing with a product form for $\mathbb{Q}[J]$,
\be
\mathbb{Q}[J]=\prod\limits_{n=1}^\infty {\cal Q}_n[\,J\,]
 \label{Qprod}
\ee
an $\ell$-point connected field correlation function can be written in view of (\ref{CWLcorrelator}) as an infinite sum over the towers,
    \begin{eqnarray}
    &&\langle\,\phi(x_1)...\phi(x_l)\,
    \rangle^{CWL}_{\rm c}=\frac{{\cal G}_{\ell}(\{ x_k \})}{\sum\limits_{n=1}^\infty\,
    n/c^l_n},           \label{1000}\\
    \nonumber\\
    &&{\cal G}_{\ell}(\{ x_k \})=\left(\frac{\hbar}i\right)^n  \left[{ \delta^{\ell} \ln\mathbb{Q}[J] \over \delta J(x_1) .. \delta J(x_{\ell})}\right] \Big|_{J = 0} \nonumber \\
    &&\qquad=\left(\frac{\hbar}i\right)^n  \sum_n \frac{\delta^{\ell} \ln Q_n}{\delta J(x_1) \cdots \delta J(x_{\ell})} \Big|_{J = 0} \qquad
    \end{eqnarray}
which we write as
\be
{\cal G}_{\ell}(\{ x_k \}) \;=\; \sum_n G_{\ell}^{(n)}(\{ x_k \})
 \label{G-Gn}
\ee
We will refer to the functions $G_{\ell}^{(n)}(\{ x_k \})$ as ``$n$-th level correlators", referring to the $n$-th level or $n$-th tower.

As we already noted in the last section, the effective couplings in this theory are $\hbar/n$ and $1/nm_p^{2}$, as opposed to $\hbar$ and $m_p^{-2}$. Thus to do a perturbation expansion it is convenient to introduce an ``effective Planck mass" given by
\be
\label{mpeff}
\tilde{M}_P(n) = \sqrt{n} m_p.
\ee
instead of the usual Planck mass. Let us assume for the time being a flat background metric $\eta_{\mu\nu}$. To do graviton perturbation expansions we will then define, for each $n$, a metric
\be
g_{\mu\nu} = \eta_{\mu\nu} + h_{\mu\nu}/\tilde{M}_P(n),
\ee
with $\tilde{M}_P(n)$ now being used instead  of $m_p$.

The action is that already used, with an Einstein action $S_G[g]$ and a matter action $S_m[\phi,g]$; writing this out explicitly we have
\begin{widetext}
\begin{align}
\label{Znfull}
\mathbb{Q}[J]  \;=\; \prod_{n=1}^{\infty} \int Dg \; D\phi_1 \cdots D\phi_n \exp\left\{{i \over \hbar} \left( n S_G[g] + \sum_{i=1}^n S_m[g,\phi_i] + \int d^4x  \frac{J(x)}{c_n} \sum_{i=1}^n \phi_i(x) \right) \right\}.
\end{align}
so that, performing the expansion in powers of $\tilde{M}_p^{-1}$, we have an $n$-th level generating functional
\begin{align}
\begin{split}
\label{Znpert}
Q_n[J] & \;\;=\;\; \int Dh \; D\phi_1 \cdots D\phi_n \exp\left\{{i \over \hbar} \left( S_G^0[h] + \sum_{i=1}^n S_m[\eta,\phi_i] \right) \right\} \\
& \qquad\qquad \times \exp\left\{{i \over \hbar} \left(  \frac{1}{c_n} \int d^4x \ J(x) \sum_{i=1}^n \phi_i(x) + \frac{1}{\tilde{M}_P(n)} \int d^4x \ h_{\mu\nu}(x) \sum_{i=1}^n T^{\mu\nu}_i(x) \right) \right\}.
\end{split}
\end{align}
\end{widetext}

We see that the matter-graviton coupling is now controlled by the effective Planck mass. Note there is a (suppressed) $\hbar$ in front of both the gravitational and matter actions, whereas the Planck mass only appears inside the Einstein-Hilbert term, so we can separately scale each of these with $n$. The matter action is evaluated in the flat background metric $\eta^{\mu\nu}$, and $S_G^0[h]$ is again the free graviton action. The Fadeev-Popov determinant has been absorbed into the $Dh$ measure, and we assume that the diffeomorphism symmetry has been sufficiently fixed to have an invertible kernel in $S_G^0$. The subscript $i$ on the stress tensors means the stress tensor of the $i$th copy of $\phi$, i.e.
\be
T_i^{\mu\nu} \;=\; \partial^{\mu} \phi_i \partial^{\nu} \phi_i - \frac{1}{2} \eta^{\mu\nu} \left[ \left( \partial \phi_i \right)^2 + m^2 \phi_i^2 \right].
\ee

The definition of the effective Planck mass $\tilde{M}_P(n)$ that we use gives the right coefficients in front of the free actions in the exponential, so that we can use Wick's theorem to write correlators as products of standard 2-point Green's functions.

In an $\ell$-point function, the $\ell$ functional derivatives with respect to the current lead to a sum of $n^\ell$ terms, ie., we have
\be
\label{correlatortotal}
G^{(n)}_{\ell}(x_1,\ldots,x_{\ell}) \;= \sum_{a_1,\ldots,a_{\ell} = 1}^{n} G^{(n)}_{a_1,\ldots,a_{\ell}}(x_1,\ldots,x_{\ell})
\ee
with the individual terms given by the connected graphs in the path integral:
\begin{widetext}
\begin{align}
\begin{split}
\label{correlator1}
G^{(n)}_{a_1,\ldots,a_{\ell}}(x_1,\ldots,x_{\ell}) & \;\;=\;\; N(n,\ell)\int Dh \; D\phi_1 \cdots D\phi_n \; \phi_{a_1}(x_1) \cdots \phi_{a_{\ell}}(x_{\ell}) \\
& \qquad\qquad \times  \left.e^{{i \over \hbar} \left( S_G^0[h] \;+\; \sum_{i=1}^n S_m[\eta,\phi_i] \;+\; \frac{1}{\tilde{M}_P(n)} \int d^4x \ h_{\mu\nu}(x) \sum_{i=1}^n T^{\mu\nu}_i(x) \right)}\,\right|_{\,\rm connected}
\end{split}
\end{align}
Here and afterwards, we define the normalization factor
\be
\label{normfactor}
N^{-1}(n,\ell) \;\;=\;\; c_n^{\ell} \int Dh \; D\phi_1 \cdots D\phi_n \; e^{{i \over \hbar} \left( S_G^0[h] \;+\; \sum_{i=1}^n S_m[\eta,\phi_i] \;+\; \frac{1}{\tilde{M}_P(n)} \int d^4x \ h_{\mu\nu}(x) \sum_{i=1}^n T^{\mu\nu}_i(x) \right)}.
\ee
and $n$ represents the CWL level/number of matter copies, $\ell$ the number of
spacetime points of the correlation function, and the $a_i = 1,...,n$ index denotes the choice of matter copy at the $i^{th}$ spacetime point in the diagram.

As is clear from \eqref{correlator1}, the correlation functions can be expanded in terms of standard QFT Feynman diagrams, and just as in standard QFT, the path integral part of this normalization factor serves to cancel the contribution of vacuum graphs - which are subtracted anyway, in the transition to the connected part. On the other hand, the $c_n^\ell$ factor in (\ref{normfactor}) is important because it reflects the modification of the formalism due to correlated world lines. To work out the correlation functions perturbatively, \eqref{correlator1} needs to be expanded order-by-order in gravitons. At lowest order, we have
\be
\label{gn0cpts}
G^{(n)}_{(0); a_1,\ldots,a_{\ell}}(x_1,\ldots,x_{\ell}) \;\;=\;\;\left. N_{(0)}(n,\ell) \int D\phi_1 \cdots D\phi_n \; \phi_{a_1}(x_1) \cdots \phi_{a_{\ell}}(x_{\ell})\; e^{{i \over \hbar} \sum_{i=1}^n S_m[\eta,\phi_i]}\,\right|_{\,\rm connected}.
\ee
where the subscripts $(r)$ on the correlator and the normalization factor indicate the order $1/M_P^r$ in the expansion in the gravitational coupling - here we are at zeroth order $r=0$.

We see the graviton path integral has factored out of both the numerator and denominator and thus cancelled out. Since $S_m$ is the \emph{free} action for a scalar field, this expression is either just the free $\phi$ propagator for $\ell = 2$ or zero otherwise, being a purely disconnected set of diagrams. Assuming $\phi \to -\phi$ symmetry, which holds in particular for a free theory, the next contribution is at order $1/M_p^2$. This is because we need to pull down two gravitons, of order $h^2$, in order to have a non-zero integral over $h$, given that the free graviton action leads to a Gaussian. Explicitly, this term  includes both the term
\begin{align}
\begin{split}
\label{gn2cpts}
& \frac{N_{(2)}(n,\ell)}{\tilde M_P^2(n)} \int Dh\; D\phi_1 \cdots D\phi_n  e^{{i \over \hbar} \left( S_G^0[h] \;+\; \sum_{i=1}^n S_m[\eta,\phi_i] \right)} \\
& \qquad\quad \left.\times \int d^4z d^4z' \sum_{i,j=1}^{n} T_i^{\mu\nu}(z) T_j^{\lambda\rho}(z')\; h_{\mu\nu}(z)h_{\lambda\rho}(z')\; \phi_{a_1}(x_1) \cdots \phi_{a_{\ell}}(x_{\ell})\,\right|_{\,\rm connected}.
\end{split}
\end{align}
\end{widetext}
which corresponds to the tadpole graph (b) and self-energy graph (c) of Fig. \ref{fig:graphs1}, along with a similar term in which the product of stress tensors $T_i^{\mu\nu}(z) T_j^{\lambda\rho}(z')$ is replaced with $\delta_{ij}\delta T_i^{\mu\nu}(z)/\delta g_{\lambda\rho}(z')$, corresponding to the seagull diagram in Fig. (d).

At this stage one sees clearly the basic intuition underlying the CWL corrections: two field copies $i \neq j$ can couple through the product of their stress tensors.

%%%%%%%%%%%%%%%%%%%%%%%%%%%%%%%%%%%%%%%%%%%%%%%%%%%%%%%%%%
\subsection{Two-point function}
 \label{G2p-prod}
%%%%%%%%%%%%%%%%%%%%%%%%%%%%%%%%%%%%%%%%%%%%%%%%%%%%%%%%%%

We begin by working out the two-point function, i.e. $\ell = 2$. The non-trivial effect here is that the field strength and mass renormalization of the $\phi$-field come in with different coefficients than those in standard QFT, essentially because one matter copy $i$ can connect to the vacuum stress tensor of another copy $j$ in a kind of tadpole diagram.

At zeroth order in gravitons, of course, the answer is unchanged from the conventional one, up to an overall normalization. Equation \eqref{correlatortotal} has $a_1,a_2$ running from $1$ to $n$, so using \eqref{gn2cpts} we get
\begin{align}
\begin{split}
G^{(n)}_{0}(x,x') & \;=\; N_{(0)}(n,2) \sum_{a,a'=1}^{n} \int D\phi_1 \cdots D\phi_n \\
& \qquad \times \; \phi_{a}(x) \phi_{a'}(x')\; e^{{i \over \hbar} \sum_{i=1}^n S_m[\eta,\phi_i]}
\end{split}
\end{align}
so that we just get $G^{(n)}_{0}(x,x') = n G(x,x')/c_n^2$, where $G(x,x')$ is the free propagator, ie., we have
\be
G(x,x') = \int d^4p \frac{e^{i p \cdot (x - x')}}{p^2 + m^2 - i \epsilon}.
\ee
The factor of $n$ appears because only the $a = a'$ elements in the sum contribute, and there are $n$ of these. The regulator  $c_n^{2}$ in the denominator comes from the definition of CWL correlators; it is the part of the normalization factor $N(n,2)$ (see \eqref{normfactor}). Summing over $n$ according to Eqs.(\ref{1000})-(\ref{G-Gn}), we have the full 2-point function to zeroth order in the Planck mass,
\be
\langle\,\phi(x)\phi(x')\,
    \rangle^{CWL}_{0} \;=\; G(x,x'),
\ee
where the coefficient $\sum n/c_n^2$ is divided out to yield the standard QFT result in the $\tilde{M}_p \to \infty$ limit.

\begin{figure}
\includegraphics[width=3.2in]{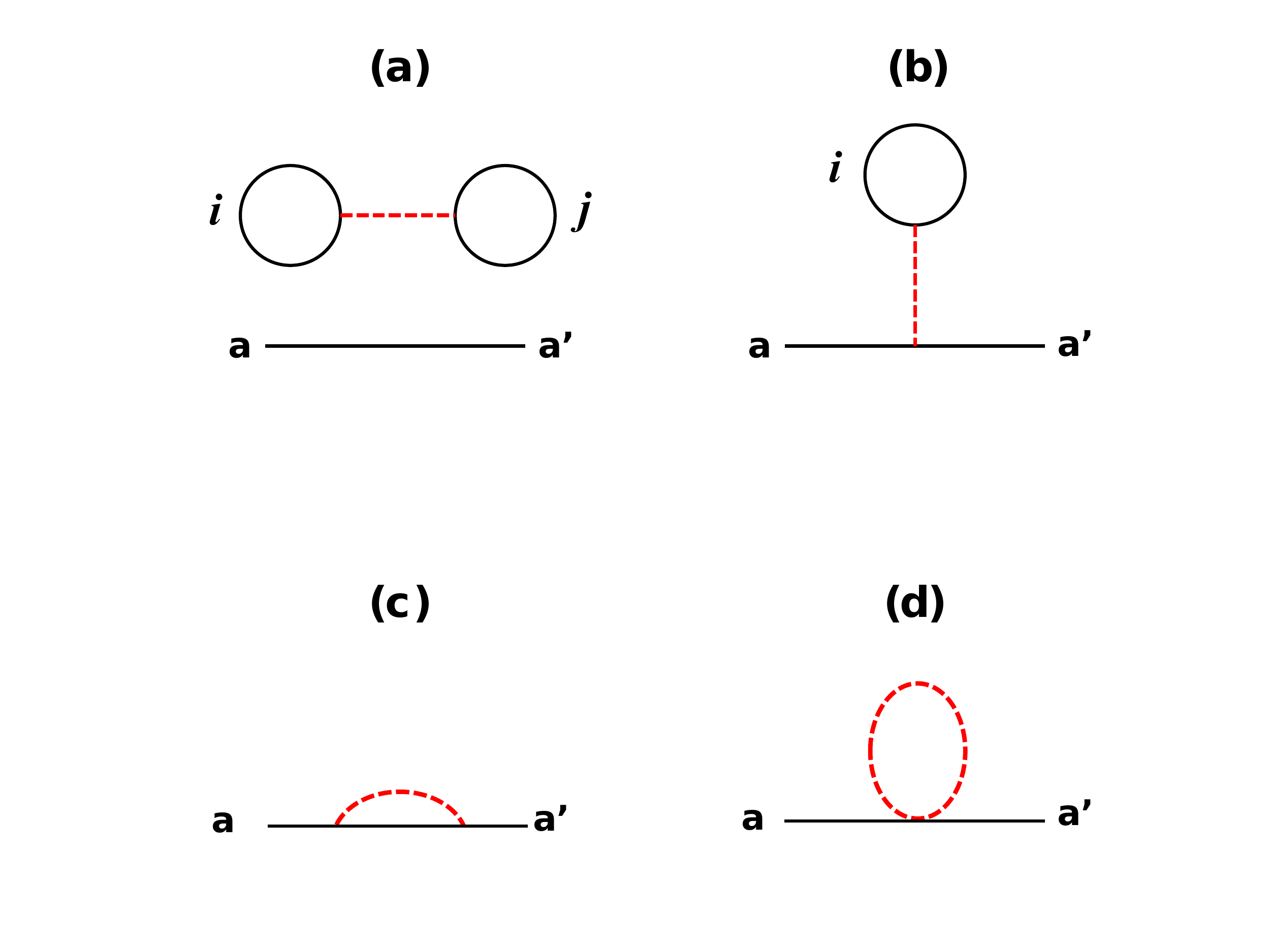}
\caption{\label{fig:graphs1} Leading order graviton terms contributing to the two-point correlator $G^{(n)}_{a,a'}(x,x')$ for product CWL theory - these corrections are calculated in the text. The four 2nd-order (in graviton insertions) graphs are (a) a disconnected graph (b) a tadpole graph (c) a 1-lopp self-energy graph, and (d) a seagull graph. For $a \neq a'$ all the diagrams vanish. For given $a=a'$, there are $n$ possible choices for the loop copy number $i$ in case (b), but only one choice in case (c). The scalar field propagator is shown in black, and the graviton propagator in red. }
\end{figure}

Much more interesting are the graviton-induced corrections; we show the leading-order contributions in Fig. \ref{fig:graphs1}, including a seagull term. The first non-vanishing correction to the two-point function comes from two graviton insertions.  Expression \eqref{gn2cpts} reads
\begin{widetext}
\begin{align}
\begin{split}
&\frac{N^{(2)}(n,2)}{\tilde{M}_P^2(n)} \int Dh D\phi_1 \cdots D\phi_n \; e^{{i \over \hbar} \left( S_G^0[h] \;+\; \sum_{i=1}^n S_m[\eta,\phi_i] \right)} \\
& \qquad\quad\times \left.\int d^4z d^4z' \sum_{i,j=1}^{n} T_i^{\mu\nu}(z) T_j^{\lambda\rho}(z') \; h_{\mu\nu}(z)h_{\lambda\rho}(z') \; \phi_{a}(x) \phi_{a'}(x')\,\right|_{\,\rm connected}.
\end{split}
\end{align}
\end{widetext}

Since the stress tensor operator $T^{\mu\nu}$ is quadratic in the matter field $\phi$, we see that unless $a = a'$ the integrals over $\phi_a$ and $\phi_{a'}$ will vanish, since the integrand will be odd under $\phi \to -\phi$. If $a=a'$, then there are three further options: zero, one, or both of the stress tensor insertions can have $i = a$. The first case, shown in Fig. \ref{fig:graphs1}(a), gives a disconnected diagram and thus cancels out. The second case, in Fig. \ref{fig:graphs1}(b), gives the $a$ propagator connected to a vacuum loop of the $i$th field. In the third case, in Fig. \ref{fig:graphs1}(c), we must have $i = j = a = a'$ and this is simply the usual self-energy diagram. In case (b), for a given external line label $a$ we have $n$ further choices of the tadpole loop, so there are $n^2$ copies of this diagram in the total correlation function at level $n$. In case (c), for given external label $a$ the internal line must also be labelled $a$, so there are only $n$ of these diagrams. By the same kind of argument the seagull graph in Fig. \ref{fig:graphs1}(d), just like the self-energy operator, enters the answer $n$ times.

Adding all these terms together, we get a $1/\tilde{M}_P^2$-contribution to the propagator
\begin{align}
\begin{split}
\label{dan2ptresult}
&{\cal G}_{(2),2}(x,x')  = \sum_{n=1}^{\infty} \frac{1}{c_n^{2}}\Big( n^2 \tilde{G}_{\rm TP}(x,x')  +  n\tilde{G}_{\rm SE+SG}(x,x')\Big) \\
& \;\;\;= \; G_{\rm TP}(x,x') \sum_{n=1}^{\infty} \frac{n}{c_n^{2}}
+ G_{\rm SE+SG}(x,x') \sum_{n=1}^{\infty} \frac{1}{c_n^{2}} .
\end{split}
\end{align}
where the tilded and untilded functions $G_{\rm TP}(x,x')$ and $G_{\rm SE+SG}(x,x')$ are the usual tadpole and 1-loop self-energy and seagull graphs in gravity theory with the respective Planck massess $m_p$ and $\tilde M_p(n)=\sqrt{n}m_p$. Note that $\tilde G_{\rm TP}(x,x')$ and $\tilde G_{\rm SE+SG}(x,x')$ are \emph{conventional Feynman
diagrams} in a theory with one scalar field coupled to gravitons,
with coupling $\tilde{M}_P(n)$. Since these both scale like $1/
\tilde{M}_P^2 = 1/n M_P^2$, we can scale out the factors of $n$,
leading to the second line of (\ref{dan2ptresult}), in which the full ${\cal G}$'s
(with no tilde) are the conventional Feynman diagrams with
coupling $m_p$ (no tilde).

%%%%%%%%%%%%%%%%%%%%%%%%%%%%%%%%%%

%%%%%%%%%%%%%%%%%%%%%%%%%%%%%%%%%%

Thus, finally in view of (\ref{1000}), we have
\be
\langle\,\phi(x)\phi(x')\,
    \rangle^{CWL}_{2}= G_{\rm TP}(x,x')+
    \frac{\sum\limits_n1/c_n^2}{\sum\limits_n n/c_n^2}
    G_{\rm SE+SG}(x,x')
\ee

If the coefficients in front of these diagrams had been the same, this would be the conventional perturbative gravity answer. It is the relative factor of $n$ in the sum that comes from CWL; this makes the tadpole graph weighted more heavily than the self-energy and seagull graphs. One can intuitively view this as each field copy's history ``gravitating'' toward the vacuum energy of the other copies.

Needless to say, all these diagrams have both UV and IR divergences, but the UV renormalization by local counterterms applies along the usual lines of BPHZ theory, though it should  be appropriately adjusted with respect to conventional QFT. Questions of renormalization for CWL theory will be discussed in future papers.

%%%%%%%%%%%%%%%%%%%%%%%%%%%%%%%%%%%%%%%%%%%%%%%%%%%%%%%%%%
\subsection{Four-point function}
%%%%%%%%%%%%%%%%%%%%%%%%%%%%%%%%%%%%%%%%%%%%%%%%%%%%%%%%%%

We now move to a four-point correlation function, that is $\ell = 4$. This behaves almost identically to the standard four-point function of perturbative quantum gravity. Equation \eqref{correlatortotal} now has $a_1, \ldots, a_4$ running from $n=1$ to $\infty$. As with the propagator above, we can expand equation \eqref{correlator1} in powers of the inverse Planck mass. The lowest-order term is zero:
\begin{widetext}
\begin{align}
%\begin{split}
G^{(n)}_{(0); a_1,\ldots,a_4}(x_1,\ldots,x_4)  \;\;=\;\; N^{(0)}(n,4) \int D\phi_1 \cdots D\phi_n \; e^{{i \over \hbar} \left( S_G^0[h] \;+\; \sum_{i=1}^n S_m[\eta,\phi_i] \right)} \; \phi_{a_1}(x_1) \cdots \phi_{a_4}(x_4)
 \;\;=\;\; 0,
%\end{split}
\end{align}
which is simply the statement that in a free theory, the four-point function consists only of disconnected diagrams. So again the interesting part is the first correction, given by
\begin{align}
\begin{split}
\label{4ptcpts}
G_n^{(2)a_1,\ldots,a_4}(x_1,\ldots,x_4) & \;\;=\;\; \frac{N^{(2)}(n,4)}{\tilde{M}^2_p(n)} \int Dh \; D\phi_1 \cdots D\phi_n \; e^{{i \over \hbar} \left( S_G^0[h] \;+\; \sum_{i=1}^n S_m[\eta,\phi_i] \right)} \\
& \qquad\qquad \times \int d^4z d^4z' \sum_{i,j=1}^{n} T_i^{\mu\nu}(z) T_j^{\lambda\rho}(z') h_{\mu\nu}(z)h_{\lambda\rho}(z')\; \phi_{a_1}(x_1) \cdots \phi_{a_4}(x_4).
\end{split}
\end{align}
\end{widetext}

Now we have four external line labels $a_1, \ldots, a_4$ to work with. Again since the stress tensor is quadratic in the matter field, this integral can only be non-vanishing if the $a_i$'s are chosen from two pairs; that is either if $a_1 = a_2 = a_3 = a_4$ or if, say, $a_1 = a_2 \neq a_3 = a_4$ or any such permutation. This leads to non-trivial combinatorics, much as in the two-point function above.

%%%%%%%%%%%%%%%%%%%%%%%%%%%%%

\begin{figure}
\includegraphics[width=3.2in]{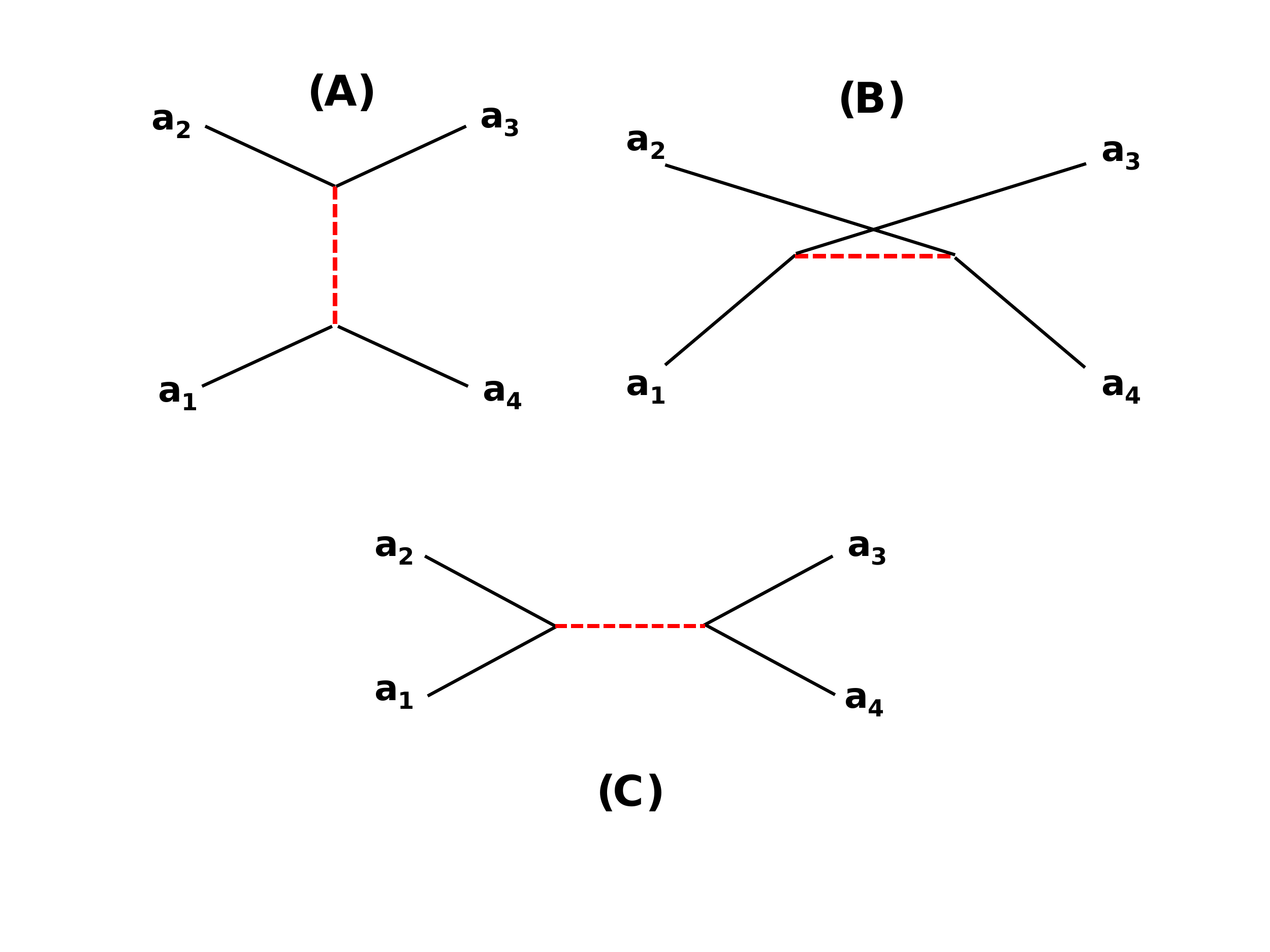}
\caption{\label{fig:graphs2} Lowest order terms contributing to the 4-point correlator $G^{(n)}_{a_1, a_2, a_3, a_4}(x_1, x_2, x_3, x_4)$ for product CWL theory - these scattering corrections are calculated in the text. The three 2nd-order (in graviton insertions) graphs are usually called direct, exchange, and crossed graphs. The scalar field propagator is shown in black, and the graviton propagator in red. }
\end{figure}

%%%%%%%%%%%%%%%%%%%%%%%%%%%%%

For the case $a_1 = a_2 = a_3 = a_4 \equiv a$, \eqref{4ptcpts} is given by the usual set of Feynman diagrams in conventional perturbative gravity coupled to a single scalar with the coupling $\tilde{M}_P(n)$. These diagrams are shown in Fig. \ref{fig:graphs2}. Note that, in contrast to the diagrams for the 2-point function considered above, we only have tree graphs for the 4-point function up to this order - thus, any results we derive should correspond to the classical scattering limit.

At CWL level $n$ there are simply $n$ of each of these diagrams 
contributing to $G_n$. However for the case of two different  
pairs, say with labels $a, a'$, we also get an additional set of  
contributions. Consider, eg., the graph in Fig. 
\ref{fig:graphs2}(a). The fields sourced at $x_1$ and $x_4$, giving the upper 2 lines in the figure, need to involve the same field  copy (ie., we require $a_2 = a_3$), and likewise for the fields sourced at $x_2$ and $x_3$. However the field copies in these two parts of the graph are in general different, ie., for $a_1 = a_4  \neq a_2 = a_3$, there are $n$ versions of the graph. Likewise in Fig. \ref{fig:graphs2}(b), for $a_1 = a_3 \neq a_2 = a_4$ there are $n$ additional versions; and in Fig. \ref{fig:graphs2}(c), for  $a_1 = a_2 \neq a_3 = a_4$ we get $n$ versions of the diagram. So in total we get $n(n-1) + n = n^2$ versions of each diagram; since the effective Planck mass $\tilde{M}_P^2 = n m_p^2$ we get an overall factor of $n$. Note the crossing symmetry between these 3 tree-level graphs.
 
However, we now notice that these combinatorial factors are
exactly the inverse of the overall normalization $N_4 = \sum_{n} n/c_n^4$.
Thus at order $1/m_p^2$,
\begin{align}
\begin{split}
\label{dan4ptresult}
\mathcal{G}_4(x_1,\ldots,x_4) & \;\;=\;\; G_4^{PQG}(x_1,\ldots,x_4).
\end{split}
\end{align}
where $G_4^{PQG}$ is the standard perturbative quantum
gravity computation at lowest order. We see that, just as for the two-point function, the 4-point CWL corrections can be evaluated entirely in terms of standard Feynman diagrams, including the full sum over CWL level $n$. However, for the 4-point function, up to this order, the CWL answer is simply the standard perturbative quantum gravity answer, with the usual Newtonian limit - quite unlike our result for the 2-point function.

A moment's thought actually makes clear the reason for this difference. We have already noted the crossing symmetry between the three 4-point graphs, yielding identical contributions to each of the s,t,u channels, so that we just get a rescaled version of the standard linear gravity 4-point function at tree level, which is then cancelled by the CWL correlation function normalization.
However, beyond tree level this argument breaks down: if we
add another graviton line in any of those diagrams to form a loop, and/or include virtual matter lines, there is no reason for the cancellation to persist. This makes it clear why there is no cancellation for the 2-point function, precisely because there are virtual matter loops, which receive extra weight with respect to diagrams without matter loops. 

In this paper we will not pursue this analysis further - a complete understanding requires a non-perturbative analysis, in which crossing symmetry is maintained.

%%%%%%%%%%%%%%%%%%%%%%%%%%%%%%%%%%%%%%%%%%%%%%%%%%%%%%%%%%%%%%%%%%%%%%%%%%%%
%%%%%%%%%%%%%%%%%%%%%%%%%%%%%%%%%%%%%%%%%%%%%%%%%%%%%%%%%%%%%%%%%%%%%%%%%%%%

\section{Summary and Conclusions}
\label{sec:summary}

%%%%%%%%%%%%%%%%%%%%%%%%%%%%%%%%%%%%%%%%%%%%%%%%%%%%%%%%%%%%%%%%%%%%%%%%%%%%
%%%%%%%%%%%%%%%%%%%%%%%%%%%%%%%%%%%%%%%%%%%%%%%%%%%%%%%%%%%%%%%%%%%%%%%%%%%%

The main goal of this paper has been to understand some aspects of the structure of CWL theories. It actually has to be counted as remarkable that it is possible to find any sort of consistent field theory which breaks some of the canonical tenets of standard QFT, and yet which reduces to ordinary QFT for low masses. In this paper we have found 2 different ways in which paths in a path integral may be correlated by gravity - the summed CWL and product CWL forms. To characterize the behaviour of these theories, we have looked at low-order perturbation theory for the propagators and field correlators; and we have looked at several consistency tests for them.

The main results of the present work have been twofold. First, we have checked the internal consistency of the two CWL theories by looking at the classical limit, at the equations of motion and Noether identities, and at the limit of small gravitational coupling. Second, we have elucidated the structure of low order perturbation expansions for both summed and product CWL.

Our main conclusions from this work have been (i) that a product form for CWL theory passes all consistency tests given here, whereas the summed form does not - it has a peculiar classical limit which is hard to understand; and (ii) that CWL correlations are generated in product CWL theory in low-order perturbation theory in the way demonstrated in the last section.

Much more detailed work is required to fully characterize product CWL theory, and show how it works in practical calculations. In particular, we need to

(i) set up both perturbative and semiclassical expansions in a consistent way, and show how both field correlators and propagators are systematically calculated and renormalized in these expansions.

(ii) derive the Ward identities in product CWL, corresponding to gauge matter transformations and to diffeomorphisms.

(iii) understand the non-perturbative aspects of the theory - particularly insofar as the crossover between quantum and classical behaviour is concerned.

(iv) calculate the observable consequences of the theory, particularly where they concern experimental deviations from quantum theory.

It turns out that these tasks can be carried through, and they are the subject of several future papers.  Beyond this there are other more technical questions that need to be clarified about non-perturbative features of the theory - most notably, how anomalies manifest themselves in the theory, and under what circumstances it is renormalizable.

%%%%%%%%%%%%%%%%%%%%%%%%%%%%%%%%%%%%%%%%%%%%%%%%%%%%%%%%%%%%%%%%%%%%%%%%%%%%
%%%%%%%%%%%%%%%%%%%%%%%%%%%%%%%%%%%%%%%%%%%%%%%%%%%%%%%%%%%%%%%%%%%%%%%%%%%%

\section{Acknowledgements}
\label{sec:ack}

%%%%%%%%%%%%%%%%%%%%%%%%%%%%%%%%%%%%%%%%%%%%%%%%%%%%%%%%%%%%%%%%%%%%%%%%%%%%
%%%%%%%%%%%%%%%%%%%%%%%%%%%%%%%%%%%%%%%%%%%%%%%%%%%%%%%%%%%%%%%%%%%%%%%%%%%%

This work has benefited from discussions with M. Aspelmeyer, H. Brown, A. Gomez, C Gooding, F. Laloe, R. Penrose, S. Reynaud, G. Semenoff, W.G. Unruh, M. Visser, and R.M. Wald, and we thank them for these.

The work was supported by NSERC in Canada, and by the Templeton foundation; one of us (AOB) also benefited from funding from the Peter Wall institute at UBC.

\end{document}